\documentclass[12pt]{article}\usepackage[hyperfootnotes=false]{hyperref}
\usepackage{epsfig}
\usepackage{amsmath}
\usepackage{amssymb}
\usepackage{graphicx}
\newlength{\abstractwidth}
\renewcommand{\thefootnote}{\fnsymbol{footnote}}
\renewcommand{\thanks}[1]{\footnote{#1}} % Use this for footnotes
\newcommand{\starttext}{
\setcounter{footnote}{0}
\renewcommand{\thefootnote}{\arabic{footnote}}}
\renewcommand{\theequation}{\thesection.\arabic{equation}}
\newcommand{\be}{\begin{equation}}
\newcommand{\bea}{\begin{eqnarray}}
\newcommand{\eea}{\end{eqnarray}}
\newcommand{\beq}{\begin{equation}}
\newcommand{\ee}{\end{equation}}
\newcommand{\eeq}{\end{equation}}

\renewcommand{\a}{\alpha}
\renewcommand{\b}{\beta}

\def\ba{\begin{eqnarray}}
\def\ea{\end{eqnarray}}

\def\12{{1 \over 2}}
\def\eq{&=&}

\def\la{\langle}
\def\ra{\rangle}

\def\simleq{\; \raise0.3ex\hbox{$<$\kern-0.75em
\raise-1.1ex\hbox{$\sim$}}\; }
\def\simgeq{\; \raise0.3ex\hbox{$>$\kern-0.75em
\raise-1.1ex\hbox{$\sim$}}\; }

\def\O2{\Omega_2}

\def\bi{\begin{itemize}}
\def\ei{\end{itemize}}

\def\sc{\setcounter{equation}{0}}

\def\W{$\Omega$}
\def\W'{$\Omega$}

\def\V{\Omega}
\def\V'{\Omega}

\def\a{{\cal{A}}}
\def\O{${\cal{O}}$}
\def\b{{\cal{B}}}

\def\c{{\cal{C}}}
\def\n{{\cal{N}}}

\def\bn{\bigskip \noindent}
\begin{document}
\renewcommand{\theequation}{\thesection.\arabic{equation}}
\begin{titlepage}
\rightline{}
\bigskip
\bigskip\bigskip\bigskip\bigskip
\bigskip
\centerline{\Large \bf {New Concepts for Old Black Holes }}
\bigskip
\begin{center}
\bf Leonard Susskind \rm

\bigskip

\bigskip

Stanford Institute for Theoretical Physics and Department of Physics, \\
Stanford University,
Stanford, CA 94305-4060, USA \\
\bigskip
\bigskip
\vspace{2cm}
\end{center}
\bigskip\bigskip
\bigskip\bigskip
\begin{abstract}
 It has been argued that the AMPS paradox implies catastrophic breakdown of the equivalence principle in the neighborhood of a black hole horizon, or even the non-existence of any spacetime at all behind the horizon. Maldacena and the author suggested a different resolution of the paradox based on the close relationship between Einstein-Rosen bridges and Einstein-Podolsky-Rosen entanglement. In this paper the new mechanisms required by the proposal are reviewed: the ER=EPR connection: precursors: timefolds: and the black hole interior as a fault-tolerant, negative information message. Along the way a model of an ADS black hole as a single long-string is explained, and used to clarify the relation between Wilson loops and precursors.

\medskip
\noindent
\end{abstract}
\end{titlepage}
\starttext \baselineskip=17.63pt \setcounter{footnote}{0}
\tableofcontents
\sc
\section{  Introduction}

My  view is that the AMPS(S) paradoxes   are quite real and imply the possibility of genuinely new phenomena behind the horizon.  But one should not be so quick to give up complementarity, which is nothing but a restatement of the equivalence principle and the unitarity of quantum mechanics. I will take it that firewalls, although possible under certain circumstances, are not inevitable, and that subtle new concepts will allow us to reconcile the existence of a conventional horizon with the principle of information conservation. The question is what new ideas have to be invoked to establish this?

The new concepts that will be reviewed and explained in this paper are: the ``ER=EPR" connection of \cite{Maldacena:2013xja}: precursor operators: timefolds: and, the black hole interior as a fault-tolerant message. None of these  are completely new; they were all implicitly or explicitly used in \cite{Maldacena:2013xja} and two of them appeared earlier in \cite{Polchinski:1999yd} and  \cite{Heemskerk:2012mn}. However they are being used in a context which is speculative with no comprehensive framework to back them up. In their defense, they are the least crazy thing I can think of.

\sc
\section{Original AMPS and ER=EPR}

\begin{figure}[h!]
\begin{center}
\includegraphics[scale=.3]{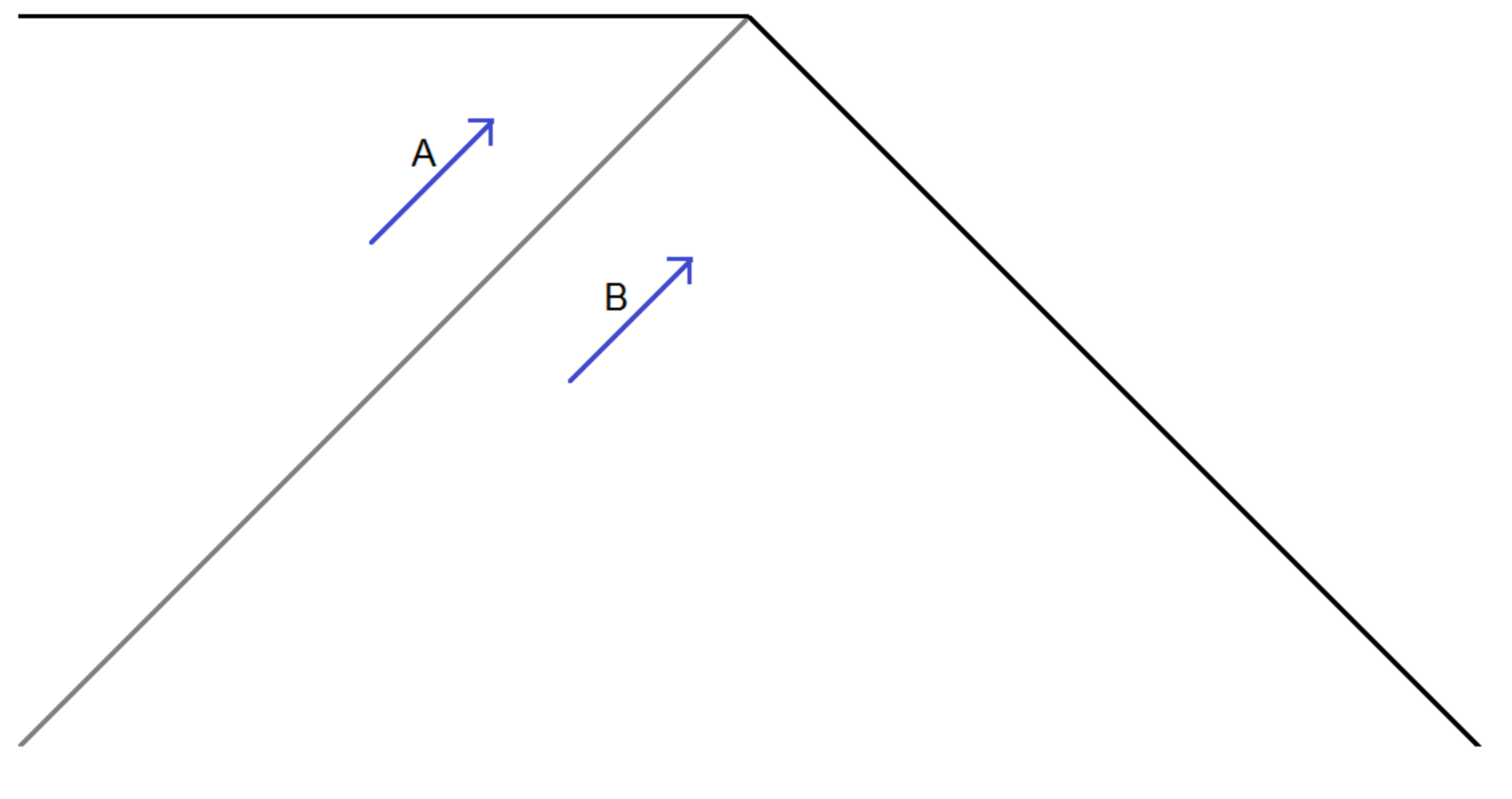}
\caption{A portion of the Penrose diagram of a Schwarzschild black hole showing a pair of entangled modes $A$ and $B.$ }
\label{AB}
\end{center}
\end{figure}

The original Braunstein-Mathur-AMPS argument \cite{Braunstein:2009my}\cite{Mathur:2009hf}\cite{Almheiri:2012rt} states that if a black hole is maximally entangled with a second distant system then the entanglement between the zone or atmosphere of the black hole, and the interior degrees of freedom of that black hole, must be disrupted. Since that entanglement is necessary for the smoothness of the horizon in the infalling frame, there must be a firewall at the horizon. The distant system could be the black hole's own radiation after the Page time; a distant gas cloud; or another black hole.
\ref{AB}).

For example let $B$ be an outgoing field-mode in the zone\footnote{The zone is the region between the stretched horizon and the photon sphere which for a Schwarzschild black hole is at $r=3MG.$} of the black hole, and let $A$ be its partner mode behind the horizon\footnote{For simplicity the modes will often be described as qubits. The three qubit operators $\sigma^+, \ \sigma^-, \ \sigma_z$ represent the creation, annihilation, and occupation-number operators.
In the notation of AMPS-AMPSS the modes $B$ and $A$ are called $b$ and $\tilde{b}$.} (See figure \ref{AB}.)
If the horizon is smooth then $B$ and $A$ must be maximally entangled. But after the Page time $B$ will be maximally entangled with a subsystem in the radiation called $R_B.$ Monogamy of entanglement precludes such double entanglement. AMPS argue that the simplest resolution of the paradox is that a firewall must replace the smooth horizon.

In the argument the radiation could be replaced by any distant system that the black hole may be highly entangled with including a second black hole.

In \cite{Maldacena:2013xja} the original AMPS argument was challenged
by a specific counterexample called the \it laboratory model. \rm
\begin{figure}[h!]
\begin{center}
\includegraphics[scale=.3]{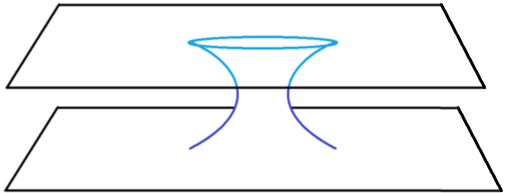}
\caption{The eternal ADS black hole is really two entangled black holes on entirely separate spaces. The entanglement is represented by and Einstein-Rosen bridge. }
\label{ERB}
\end{center}
\end{figure}
The laboratory model is essentially the eternal two-sided ADS black hole \cite{Maldacena:2001kr} describing two maximally entangled black holes on entirely disconnected spaces (See figure \ref{ERB}). For the sake of definiteness, when it matters I will assume that the size of the black holes is the ADS radius, the entropy is $N^2,$ and the temperature equals the inverse ADS scale. Equivalently, the system is just above the Hawking-Page transition.

\begin{figure}[h!]
\begin{center}
\includegraphics[scale=.3]{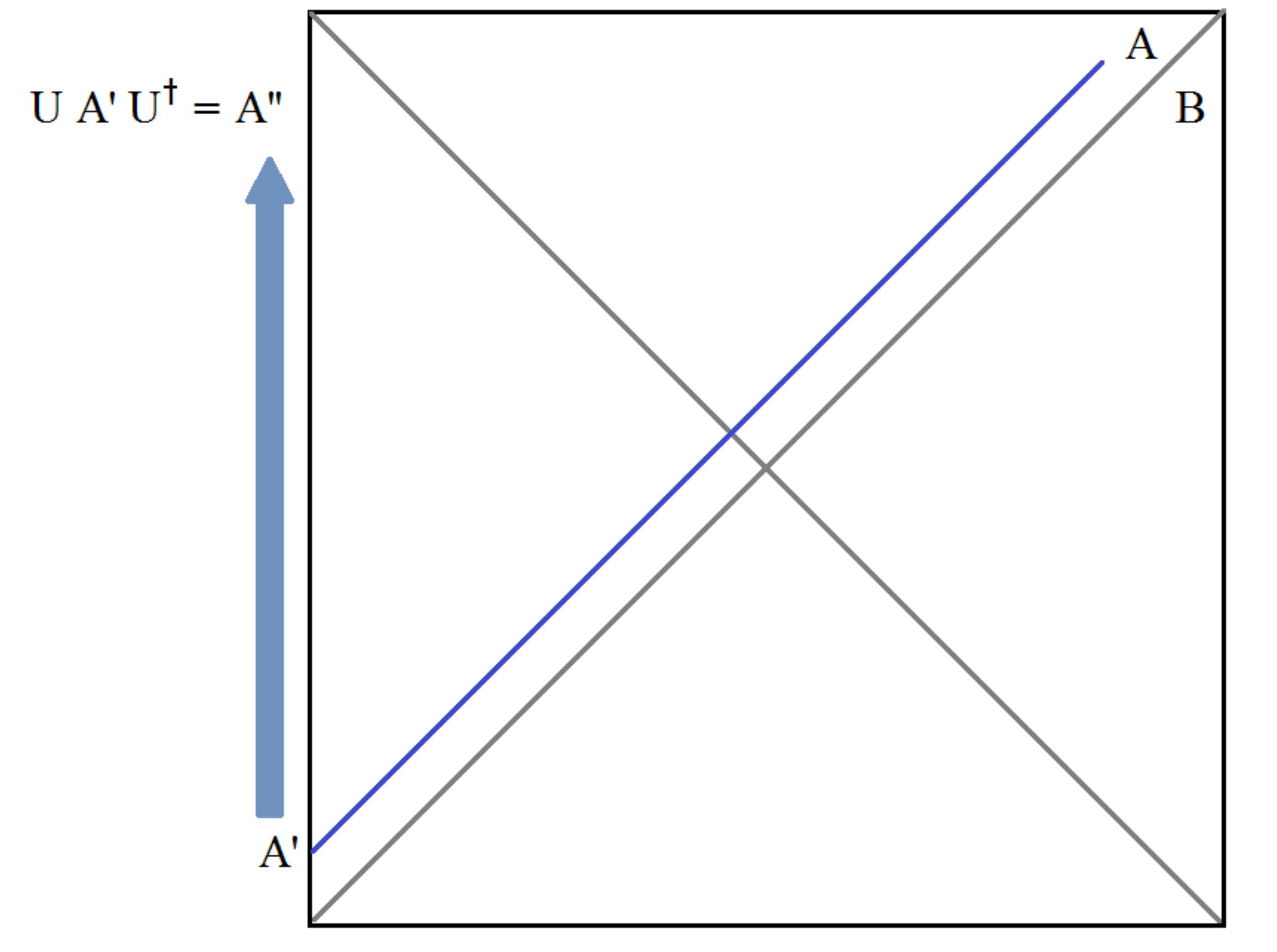}
\caption{ The eternal black hole is really two entangled but non-interacting black holes. The black hole on the right side has paired exterior and interior modes $B, \ A.$ A signal can be sent from $A'$ to $A.$ The operator $A''$ is a very non-local precursor obtained by time evolution from $A'.$}
\label{eternal}
\end{center}
\end{figure}

The two black holes can be thought of as being separated by an infinite distance so that they do not interact at all. One may think of one side---call it the right side---as containing the black hole that Bob has access to. The left side represents the distant system that Bob's black hole is entangled with. In other words it replaces the early Hawking radiation in the AMPS argument. The Penrose diagram for the eternal black hole is shown in figure \ref{eternal}.

The fact that the Penrose diagram is connected by an Einstein-Rosen bridge does not represent either interaction or a finite separation between the black holes. It represents the entanglement between them \cite{VanRaamsdonk:2010pw}.

Alice controls the left CFT. I'll assume that she has a quantum computer which is powerful enough to apply any unitary operator, no matter how complicated, to the left CFT in an arbitrarily short time. Alice's powers may seem excessive \cite{Harlow:2013tf}, but I will play by the rules of AMPS in which any experiment is possible so long as it does not conflict with the principles of quantum mechanics.

At time $t=0$ the two CFT's are set up in the thermofield-double state describing two entangled black holes.
It is generally agreed that such black holes have smooth horizons. Thus despite satisfying the conditions for the AMPS argument, there are no firewalls
 for this system. One thing is clear from this example: the assumption that the interior of a black hole must be constructed from degrees of freedom physically close to the black hole cannot be correct \cite{Marolf:2012xe}. In \cite{Susskind:2013tg} it was pointed out that this ``proximity postulate" was an underlying assumption of the AMPS paradox. In the laboratory model the proximity postulate is maximally violated; the construction of the interior of Bob's black hole is in terms of degrees of freedom that, at least in part, are in the infinitely distant left side.

\begin{figure}[h!!]
\begin{center}
\includegraphics[scale=.3]{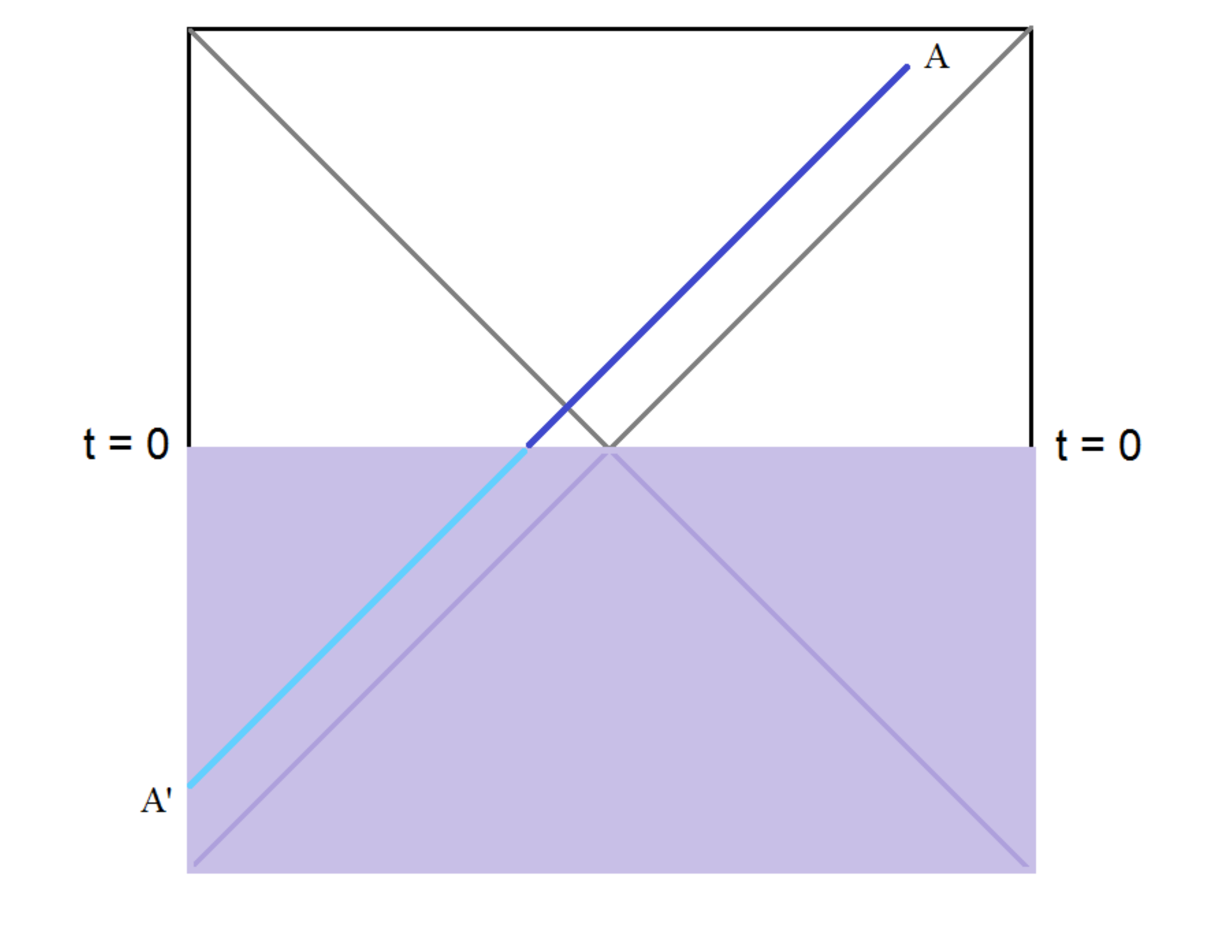}
\caption{The earlier half of the eternal black hole history can be thought of as fictitious. The black hole is created at time $0$ in the state that would have resulted from the earlier history. If there is no operator $A'$ activated in the fictitious past then the state is the Thermofield double state. }
\label{fiction}
\end{center}
\end{figure}

The geometry shown in figure \ref{eternal} is eternal-to-the-past as well as to the future. For our purposes we will assume that the lower half of the diagram is a fiction, useful for describing the way that Alice set up the initial state at $t=0.$ This is depicted in figure \ref{fiction}.
The initial state associated with figure \ref{fiction} without the blue diagonal line is the thermofield-double state or equivalently, the Hartle Hawking state.

If an operator such a $A'$ acts before $t=0$ \cite{VanRaamsdonk:2010pw}\cite{Shenker:2013pqa} then it represents a particular perturbation on the initial state in the bulk. For the most part we can just pretend the earlier half of the diagram exists, but in some contexts it will be important to consider it fictional.
By contrast with $A',$ the operator $A''$ is defined in the future of $t=0$ and represents something that Alice can actually measure.

It was suggested in \cite{Maldacena:2013xja} that the mechanism for encoding the interior of one ADS black hole in terms of its distant entangled partner is similar to the $A=R_B$ proposal for a very old evaporating black hole \cite{Susskind:2013tg} (Warning! see however Section 2.2).
$A=R_B$ gets around the double entanglement by postulating that the interior degrees of freedom are encoded in a subsystem $R_B$ of the radiation degrees of freedom. If $A=R_B$ then there is no double entanglement: $B$ is entangled with both $A$ and $R_B$ because $A$ and $R_B$ are the same thing. As we will see in section 2.2 the identification of $A$ and $R_B$ is not consistent and will lead to serious contradictions . Nevertheless let's follow the logic of  $A=R_B.$

The AMPS argument against $A=R_B$ is very simple. Alice measuring\footnote{The terms \it measuring \rm and \it applying \rm an operator have different meanings. For simplicity suppose the operator $Q$ is a component of a qubit and is therefore both hermitian and unitary. For Alice to apply $Q$ means that she interacts with the system so that
$$
|\Psi\ra \to Q|\Psi\ra.
$$
To measure $Q$ she needs a second apparatus-system that becomes entangled with $Q.$ For example
$$
|\Psi,0\ra  \to
\frac{1+Q}{2}|\Psi,1\ra +\frac{1-Q}{2}|\Psi,2\ra
$$
where $|0\ra, $  $|1\ra,$ and  $|2\ra,$ are states of the apparatus.  For most purposes in this paper the difference is not important.}
 a component of the qubit $R_B$ will necessarily disturb the mode $A$ and create a particle just behind the black hole horizon. Bob falling into the black hole will be hit by the particle. That in itself is not a problem; Alice simply did something that created a particle in a particular mode, and Bob detected it. ( Incidently, this example illustrates the importance for AMPS of assuming that Alice can carry out arbitrarily complex operations---measuring $R_B $ is expected to be extremely difficult \cite{Harlow:2013tf}.)

But AMPS argues that Alice's experiment, which took place in the extremely distant early radiation, could not possibly have affected the black hole interior. It was too far away. Thus the particle that Bob experiences must have been there whether or not Alice acted on $R_B.$ By the same argument there must have been a particle in every mode of the zone. Those particles comprise the firewall.

The mechanism that \cite{Maldacena:2013xja} invoked to  refute this argument in the laboratory model is ER=EPR \cite{Maldacena:2013xja}. Despite being infinitely separated in external space, as a consequence of the Einstein-Podolsky-Rosen entanglement the two black holes are connected by an Einstein-Rosen bridge. If Alice disturbs the second black hole a particle may indeed be sent to the interior of Bob's black hole; not through the external space, but through the Einstein-Rosen bridge as in figure \ref{eternal}.
Similarly, according to \cite{Maldacena:2013xja},  ER=EPR asserts that a black hole is connected to its own radiation by a bridge and that disturbing $R_B$ sends a particle to Bob through the bridge.

From inspection of figure \ref{eternal} we see that there is a way of sending a signal from the left side at $A'$ to the region just behind Bob's horizon. By activating an operator at $A'$ Alice can populate the mode $A$ so that Bob experiences a particle in that mode. That is not surprising since $A'$ is in the causal past of $A.$

The operator $A'$ is not the analog of $R_B$ for the two sided laboratory case. The analog of $R_B$ is an operator $A''$ in the left-side CFT, but which Alice can measure at a much later time than $A'$. $A''$ is defined in such a way that it has the same effect as $A'.$ More precisely
$A''$ is defined so that if it acts at a certain later time (than $A'$) its action is to produce exactly the same final state as the action of $A'$ would produce, if it acted at the earlier time. We define the left and right side time evolution operators by

\bea
U_L \eq e^{-iH_L t} \cr \cr
U_R \eq e^{-iH_R t}
\eea

\bigskip
\noindent
where $t\equiv (t''-t')$ is the time between $A'$ and $A''$ and $H_L, \ H_R$  are the left and right CFT Hamiltonians. We can define $A''$ by,

\be
A''  = U_L A' U_L^{\dag}
\label{A'' = uA'udag}
\ee

\bigskip
\noindent
The role of $A''$ is similar to the role of $R_B$ in the evaporating case.
If Alice measuring or acting with $A''$  creates a particle at $A,$  it
does not follow that the particle would have been there if Alice had not acted. The hypothetical particle was created by Alice, and arrived at $A$ through the Einstein-Rosen bridge.

For the moment we will accept that Alice's measurement of $A''$ does in fact create such a particle, but later in Sections 2.7 and 4.9 arguments will be made  that a  new kind of  ambiguity manifests itself in quantum gravity.

Note that nothing in the above violates the non-traversability of the Einstein-Rosen bridge; Bob can only receive the particle after he has crossed the horizon of his black hole. Non-traversability is absolutely guaranteed by the fact that the two CFT's do not interact.

\subsection{Does $A=R_B \ ?$}
Although it doesn't matter for the argument above, $A=R_B$ must not be taken as an operator equivalence. All sorts of mischief will follow from taking it literally.
The same is true of $A=A''.$  We will concentrate on the laboratory case but the evaporating black hole is similar.

Assume for the moment that $A=A''$ is an operator-identity relating a bulk operator $A$ associated with Bob's black hole, to a left-side CFT operator $A''$. Suppose that Bob throws some particles into his black hole which were aimed to collide at $A$ and perturb it. If $A=A''$ is an operator identity, Bob would have succeeded in sending a signal to the left side, from outside his black hole on the right side. This violates the non-traversable property of Einstein-Rosen bridges.

$A''$ is just $A'$ written in terms of operators at a later time, so one can ask if the identification of $A$ with $A'$ is universally correct. It is not hard to see why it cannot be. Consider the propagation of a wave from $A'$ to $A.$ It passes through the black hole geometry on both sides and in the process will get scattered. Some of the wave from $A'$ will get scattered back to the left boundary. This is enough to preclude an exact identification.

More to the point, the fact that Bob, on the right side, can send in particles which disturb $A$ means that it has dependence on the right side as well as the left side. From the discovery of a particle in the $A$ mode one cannot conclude that it came from $A';$ it may have originated from matter that fell in from the right side. A formal identification of $A$ with $R_B$ does not make sense.

Nevertheless the idea that by measuring $A''$ Alice can send a signal to the interior of Bob's black hole seems correct. The Penrose diagram \ref{eternal} for the Thermofield Double state, obviously allows a particle to be sent from $A'$ to $A$ with significant probability. Since operating with $A''$ at the later time has the same effect as operating with $A'$ at the earlier time, it follows that measuring or acting with $A''$ can send a signal to $A.$ In \cite{Maldacena:2013xja} instead of writing $R_B = A,$  $A''= A$ or $A'=A,$ the notation

\bea
 R_B & \longmapsto & A \cr \cr
 A'' & \longmapsto & A  \cr  \cr
  A'' & \longmapsto & A
\eea

\bigskip
\noindent
was advocated, indicating that the left side determines $A$ through bulk evolution.

\bigskip

\bigskip

Two important things can be seen from this example.
\bi
\item The first part of the AMPS argument is correct: measuring or disturbing $R_B$ or $A''$ can send a disturbance to Bob just behind the horizon of his black hole.
\item The second part is wrong. There is no implication that the particle would appear at $A$ if Alice had not acted with $A''.$ Therefore there is no need for a firewall.
\ei

To understanding the ER=EPR connection better we need to know more about the mechanism that allows Alice to send her message. This involves two somewhat unfamiliar concepts: precursors and timefolds.

\subsection{Precursors and Timefolds}
Precursors were introduced in \cite{Polchinski:1999yd} as highly non-local CFT operators that contain information about events which eventually will manifest themselves as local boundary excitations. More generally any non-local operator which acts at one time, to simulate the effect of a local operator acting at a different time (later or earlier), will be referred to as a precursor.

Consider a quantum system with Hamiltonian $H.$ We will work in the Schrodinger picture. Suppose the evolution of the system is perturbed by applying an operator $X$ at time $t=0.$ Let us ask if there is an operator that we can apply at a later time $t>0$ which gives rise to the same state that would have resulted from applying $X$ at $t=0.$ For example suppose that Alice had intended to apply $X$ at $t=0,$ but forgot to do so. She wishes to correct her mistake by applying $Y$ at a later time $t.$ What operator should she apply? The answer is the precursor\footnote{I originally planned to call $Y$ a post-cursor in this case since it acts later than $X.$ However, in the interests of minimizing new terminology I will call all such operators precursors, regardless of the sign of $t.$} $Y$ defined by,

\be
Y = e^{-iHt} X e^{iHt}
\ee

\bigskip
\noindent
or using the notation $U(t) = e^{-iHt},$

\be
Y = U(t) X U^{\dag}(t)
\ee

\begin{figure}[h!]
\begin{center}
\includegraphics[scale=.3]{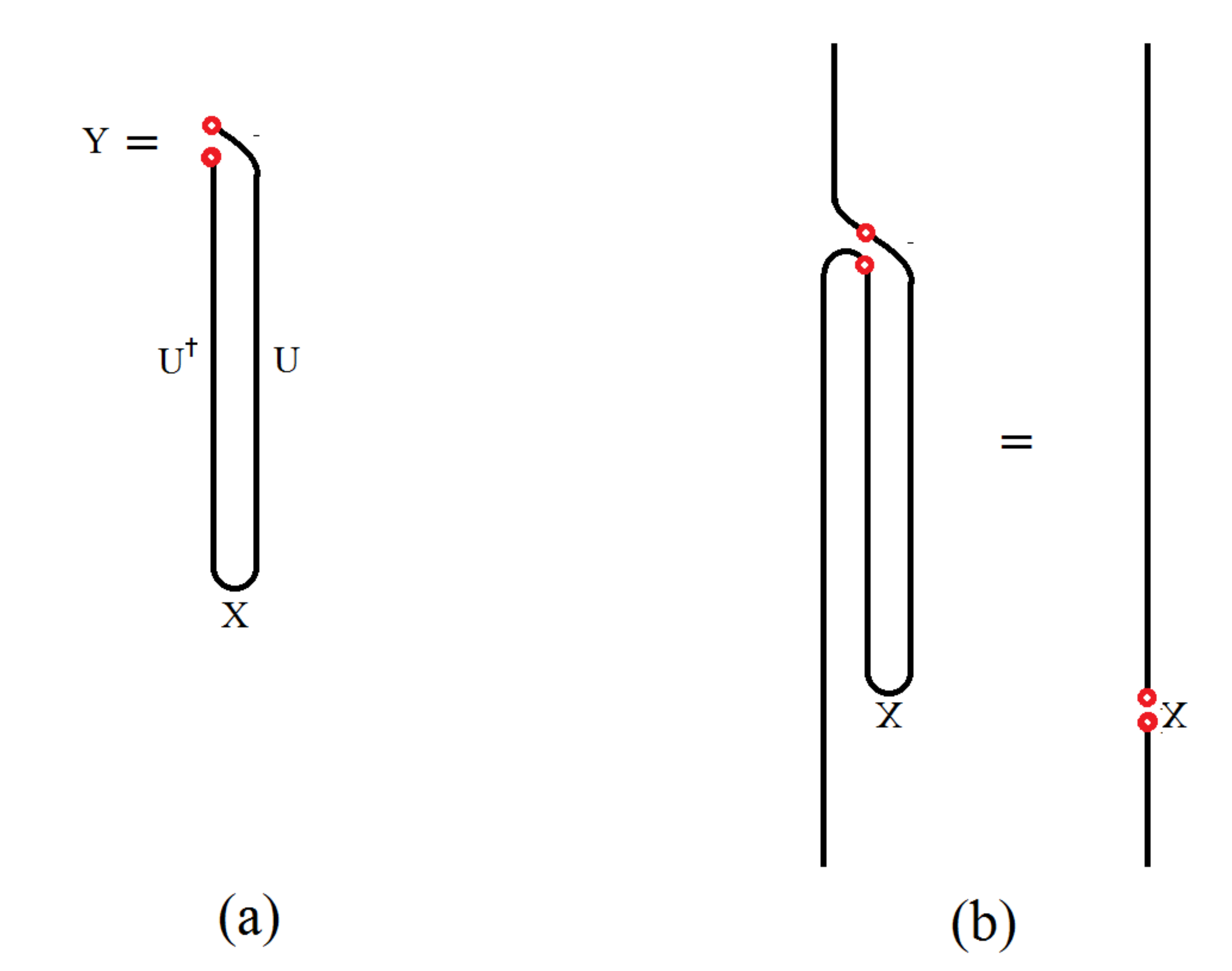}
\caption{A precursor $Y$ is constructed by propagating back to an earlier local operator $X.$ The diagram is a timefold. If the precursor is applied at one time the later effect is the same as if $X$ had acted at the earlier time. }
\label{hairpin}
\end{center}
\end{figure}

\bigskip
\noindent
Pictorially we can represent the insertion of $Y$ as in figure \ref{hairpin}.

The operator $Y$ is called a precursor of $X$ and the diagram representing it in figure \ref{hairpin} is a timefold.
We can consider a very simple example of a free one-dimensional particle with Hamiltonian $H=p^2/2.$ Let $x$ represent the position of the particle. The precursor of $x$ is

\be
y=e^{-i\frac{p^2 }{2}t} \ x \ e^{i\frac{p^2 }{2}t} = x - pt
\ee

\bigskip
\noindent
Generally calculating a precursor of a given operator is computationally extremely difficult, especially if the system is chaotic and the time interval $t$ is comparable to the scrambling time \cite{Sekino:2008he} or longer. As an example consider a box of interacting particles and let $X$ represent the position of one of the particles. After a short period of time during which the particle scatters a few times the precursor becomes much too complicated to write down. But nevertheless it is well defined.

The action of precursors can be described by a path integral procedure that involves a time-fold. Suppose the initial state is given at a time $t_i$ and final state at $t_f. $ The precursor acts at time $t$ with $t_i<t<t_f.$

\be
|\psi_f\ra = U(t-t_i) \ Y \ U(t_f -t) |\psi_i\ra
\ee

\bigskip
\noindent
We may replace this by

\bea
|\psi_f\ra \eq U(t-t_i) \ U(t) X U^{\dag}(t) \ U(t_f -t) |\psi_i\ra \cr \cr
\eq U(t-t_i) \ U(t) X U(-t) \ U(t_f -t) |\psi_i\ra
\eea

\begin{figure}[h!]
\begin{center}
\includegraphics[scale=.3]{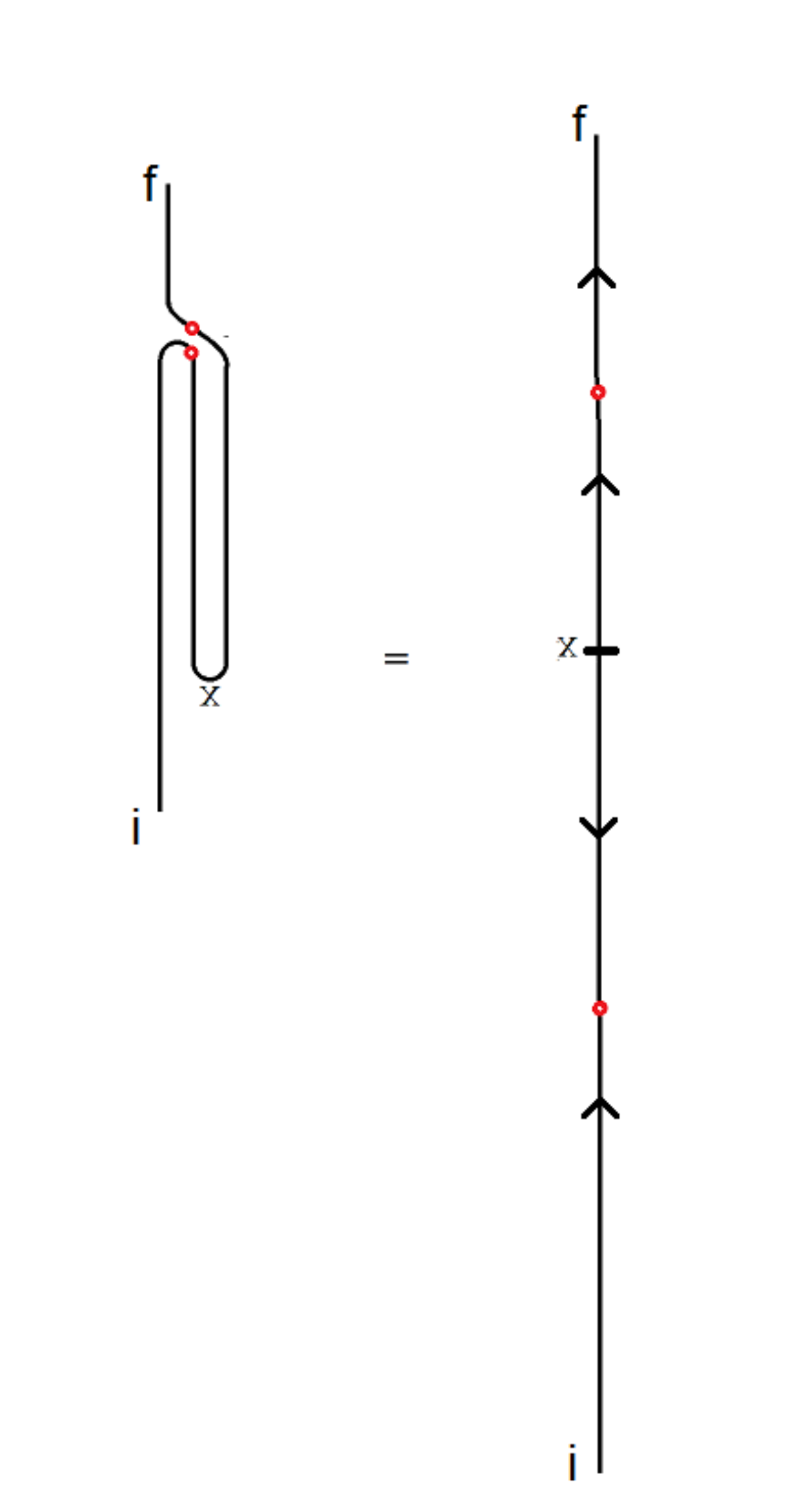}
\caption{ Precursors can be calculated by path integration. During the portion of the path with the downward arrow the sign of the Lagrangian is changed.}
\label{path}
\end{center}
\end{figure}

\bigskip
\noindent
The entire evolution can be described by a single path integral where,
during the downward
part of the evolution the Hamiltonian is reversed in sign as in figure \ref{path}.
The periods indicated by upward arrows are governed by the usual Hamiltonian, but during the interval with a downward arrow the Hamiltonian has the opposite sign.

To give a graphic example consider a closed system containing Professor Strangebody's cat. Strangebody has insisted that his assistant Miss Goody kill the cat by shooting a bullet (acting with the unitary operator $X$). Goody fools Strangebody by waiting a few years---longer than the cat's lifespan---and then acts with the precursor $Y$. When Strangebody later looks in the box he congratulates Goody for carrying out the mission successfully. The cat of course did not experience the unpleasantness during her lifetime. Or so Goody believes.
We will see that the truth is much  more subtle.

\subsection{Timefolds and Bulk Geometry}

Let's use a timefold to define a CFT operator (in a one-sided theory) that can act at $t=0$ to create a particle in the bulk of ADS. Such an operator clearly has to be a non-local CFT construction. Let $\cal{O}$ be a local boundary operator dual to some bulk operator such as the dilaton. The precursor

\be
U(t) { \cal{O} } U^{\dag}(t)
\label{O-precursor}
\ee

\begin{figure}[h!]
\begin{center}
\includegraphics[scale=.3]{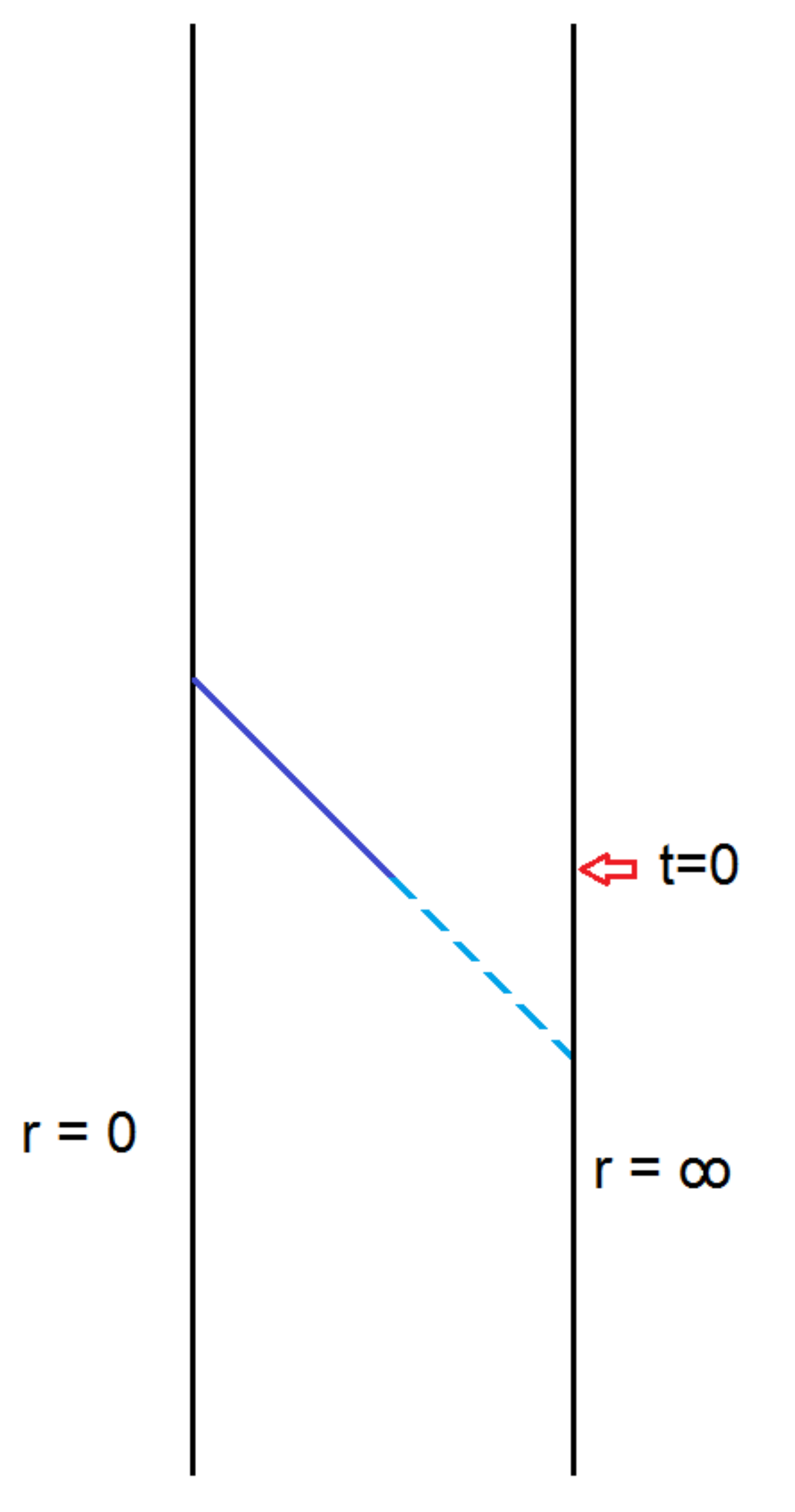}
\caption{A boundary source acts at time $-t$ and creates a particle. At time $0$ the particle is in the bulk. The precursor, indicated by the red arrow, creates the particle already in the bulk, but the point of creation is ambiguous. }
\label{create1}
\end{center}
\end{figure}

\bigskip
\noindent
acting at $t=0$ creates the same state as acting with $\cal{O}$ at time $-t.$ It creates a dilaton in the bulk, propagating inward from the boundary, as shown in figure \ref{create1}.

The question is exactly when the particle is created; in other words, on what spacelike surface does the particle materialize when \ref{O-precursor} acts? The precursor acted in the CFT at $t=0$ but that does not uniquely specify a time-slice through the bulk. To understand this better we recall that the bulk dual of a CFT wave function at time $t$ is a Wheeler DeWitt wave function. The WDW wave function describes the history of a region consisting of all spacelike surfaces that end on the boundary at the time $t.$

\begin{figure}[h!]
\begin{center}
\includegraphics[scale=.3]{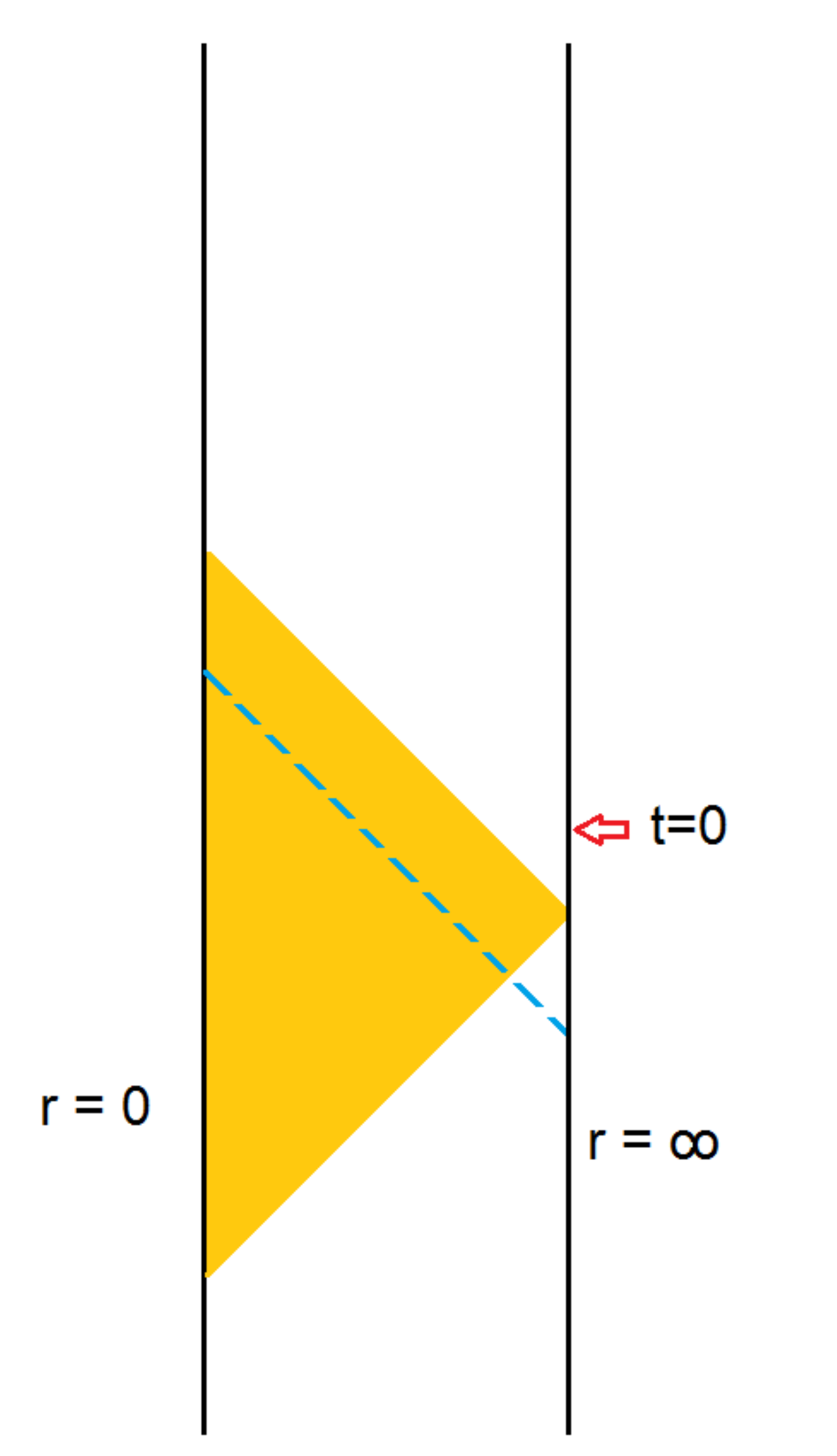}
\caption{The WDW history for a slightly negative time is shown in yellow. Since the precursor has not yet acted the history must be consistent with no particle. }
\label{create2}
\end{center}
\end{figure}

Now consider the WDW history associated with a wave function at a slightly negative time as in figure \ref{create2}.
Since the precursor has not acted yet, the wave function is exactly the same as it would have been if no particle had been injected. Therefore the history must not contain the particle.

\begin{figure}[h!]
\begin{center}
\includegraphics[scale=.3]{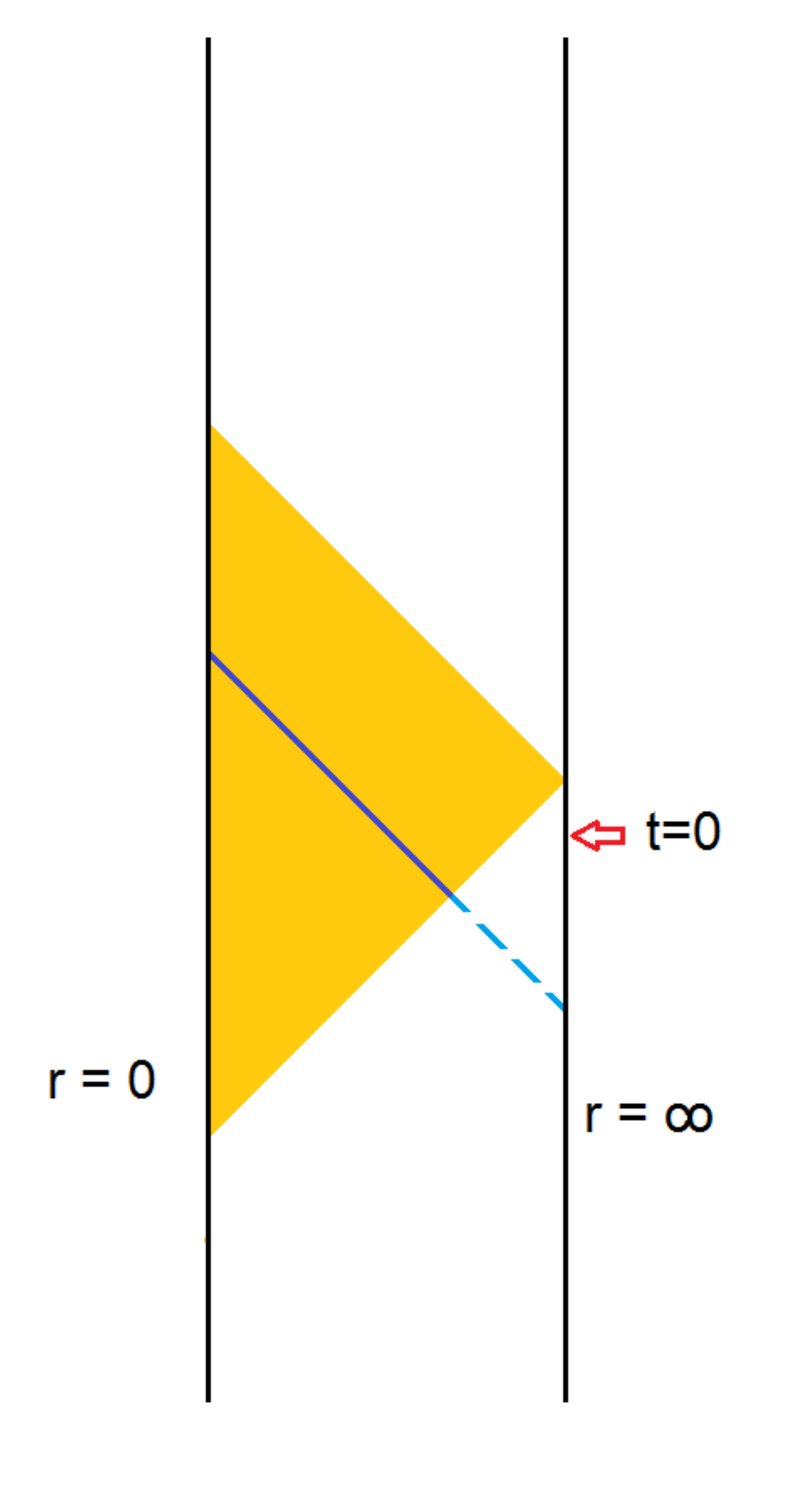}
\caption{The WDW history for a slightly positive time. The precursor has acted so the history must contain the particle. }
\label{create3}
\end{center}
\end{figure}

\begin{figure}[h!]
\begin{center}
\includegraphics[scale=.3]{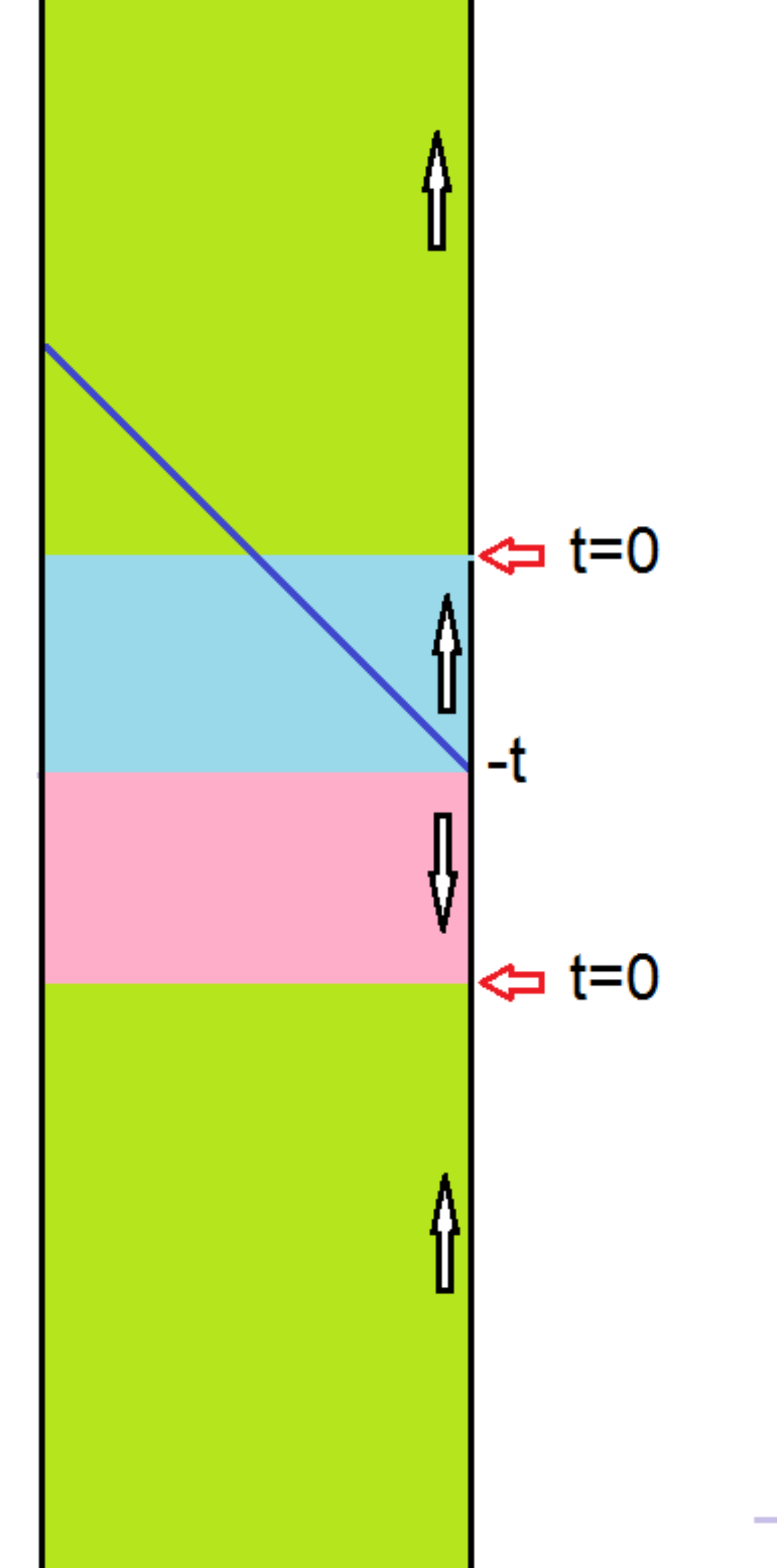}
\caption{In the timefold picture the particle is uniquely created on the boundary at a specific time in the fold. The blue and pink regions are the two pieces of the fold. In the pink region the time-time component of the vierbein changes sign.}
\label{create4}
\end{center}
\end{figure}

On the other hand, if we consider the wave function for slightly positive time, the precursor would have acted, and the corresponding  history must contain the particle. This is shown in figure \ref{create3}.
The problem is obvious. The two histories do not agree in the overlap region. We can either accept a degree of ambiguity in when the particle was created, or we can can view the process as a folded geometry (figure \ref{create4}).
\begin{figure}[h!]
\begin{center}
\includegraphics[scale=.6]{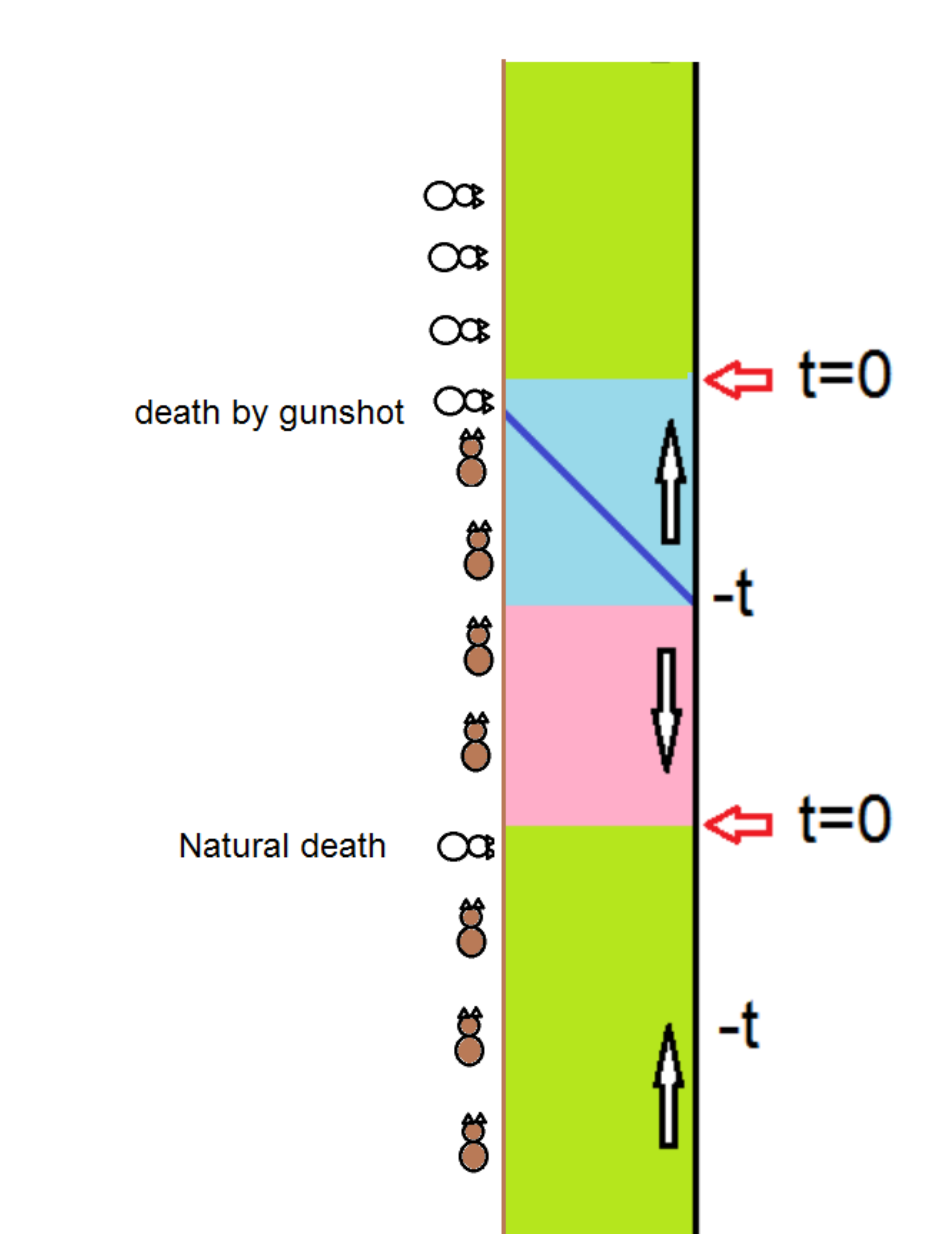}
\caption{Strangebody's experiment. The brown vertical cats are alive and the horizontal white ones are dead. The blue line represents the bullet in the timefold region.}
\label{natural}
\end{center}
\end{figure}
In the folded description the particle is created at a specific point on the boundary along the fold. If one accepts the fold as a legitimate bulk geometry there is no ambiguity about where it is created. Notice that all times between $-t$ and $0$
are represented by three separate instants in the timefold.

In the ADS-CFT context the timefold can be defined by altering the boundary Hamiltonian in a time dependent way. Suppose the usual Hamiltonian of the CFT is $H.$ Multiplying $H$ by a positive function $f(t)$  is equivalent to a coordinate change $t\to t'$ where $$f(t) = \frac{d t'}{d t}.$$
It may also be defined by altering the boundary conditions on the metric, or more conveniently, the time-time component of the vierbein $e^0_0$. The timefold is a generalization defined by allowing $f(t) $ and $e^0_0$ to become negative. For example they may be chosen to be $-1$ over some interval. This is equivalent to changing the sign of the boundary CFT hamiltonian over the same interval.

How the change of Hamiltonian extrapolates to a change in the bulk geometry is presumably determined by bulk equations. For a geometry that  has a time-like Killing vector  it is just a matter of extrapolating the timefold along  constant time surfaces orthogonal to the Killing vector. During the first half of the fold $e^0_0$ is negative, and then  flips sign in the second half of the fold.

Now let's return to Strangebody and Goody. Drawn as a timefold  the history is more complicated then what Goody thought.
Figure \ref{natural} shows the folded history of Strangbody's experiment. Initially, the cat was alive until it died a natural death just before Goody acted with the precursor at $t=0.$ Strangebody had instructed Goody to shoot at time $-t,$ but she didn't do it. However inside the time-folded region the cat is resurrected and then shot before the time-fold ends.

Acting with the precursor created two versions of history, one in which the cat died of natural causes, and one in which it was shot. Notice that this ambiguity has nothing to do with branching of the wave function of the many-world type. This type of ambiguity seems to be the cause of some of the confusion about black hole interiors.

\subsection{Timefolds and Alice's Signal to Bob}

The simplest example of a timefold is to insert the unit operator into a time evolution in the form $U(t)U(-t)$ or $U(t) U^{\dag}(t).$
In ordinary quantum mechanics this is a completely trivial operation.
However, timefolds become interesting when there are dual descriptions of the same quantum system. In that case one half of the timefold may be described in one description and the other in the dual description.
They are particularly interesting in situations with horizons. For example the excursion into the past may be expressed in the boundary-language of the dual CFT, and the return evolution in the language of the infalling frame.

Let us return to the laboratory model and the operators $A,$ $A',$ and $A''.$ It is obvious from figure \ref{eternal} that a signal can be sent from $A'$ to $A$ through the Einstein-Rosen bridge. This is unsurprising since $A'$ is in the causal past of $A. $

\begin{figure}[h!]
\begin{center}
\includegraphics[scale=.3]{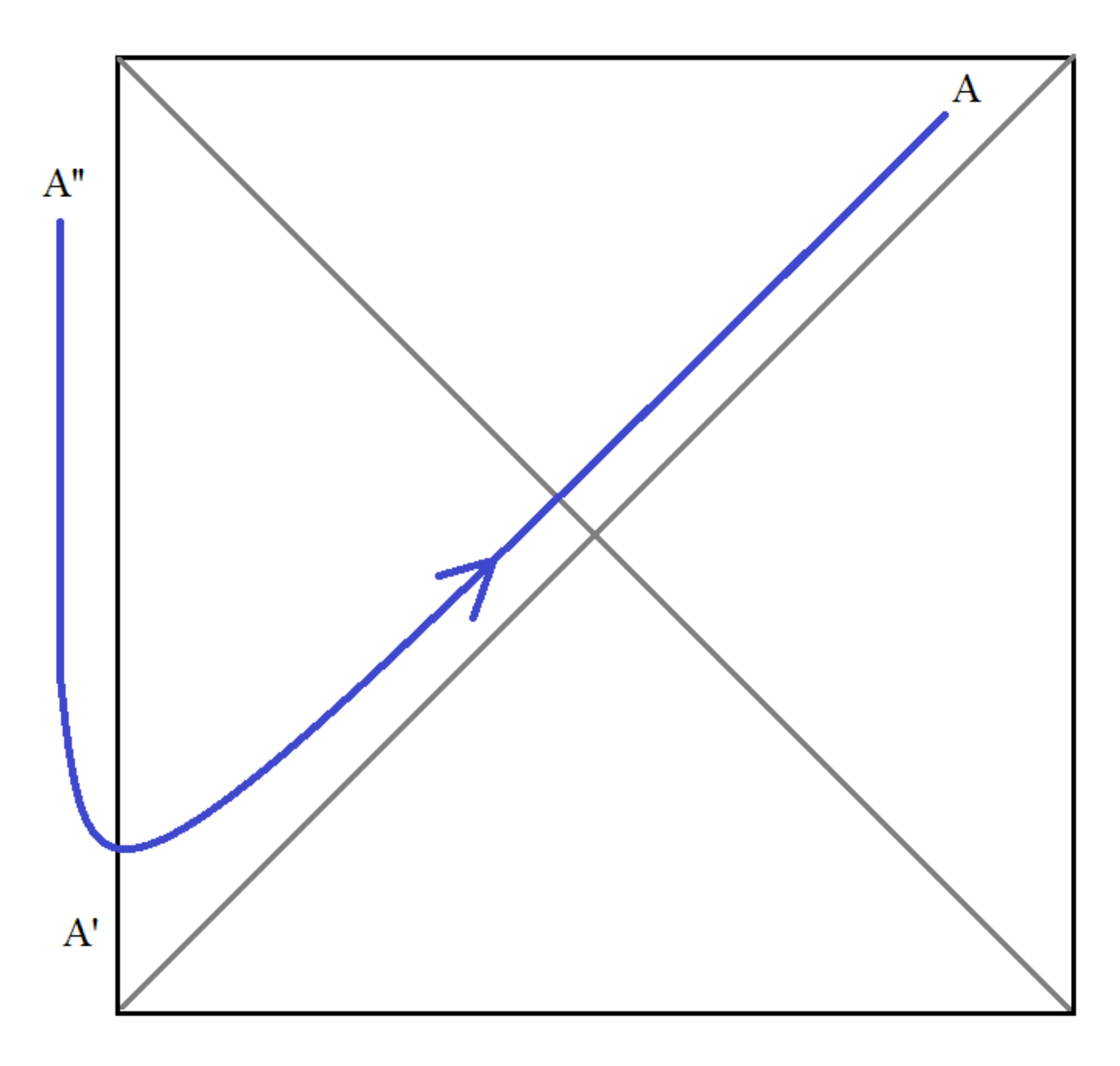}
\caption{The timefold from $A''$ to $A$ and back, can be described in two halves; one in the boundary theory and one in the bulk theory. }
\label{whoosh}
\end{center}
\end{figure}

On the other hand sending a signal from $A''$ to $A$ looks very doubtful if one supposes that $A''$ is a localized operator with support in the upper left corner of the diagram. However, from \ref{A'' = uA'udag} one sees that $A''$ is a precursor obtained from $A'$ by a timefold. It is constructed so that if it acts on the left CFT at the later time in figure \ref{eternal}, the effect is identical to that of $A'$ at the earlier time. A useful way to think about it is to envision
the first half of the precursor---the past-directed half--- as occurring in the left-boundary CFT: the second future-directed half as occurring in the bulk, and sending the signal to $A$ as in  figure \ref{whoosh}.
Thus the effect of Alice operating with $A''$ is the same as if she had (but didn't ) apply $A'$ earlier. Figure \ref{eternal} is misleading because $A''$ is very far from being a local CFT operator or a bulk operator localized near the left boundary.

\begin{figure}[h!]
\begin{center}
\includegraphics[scale=.3]{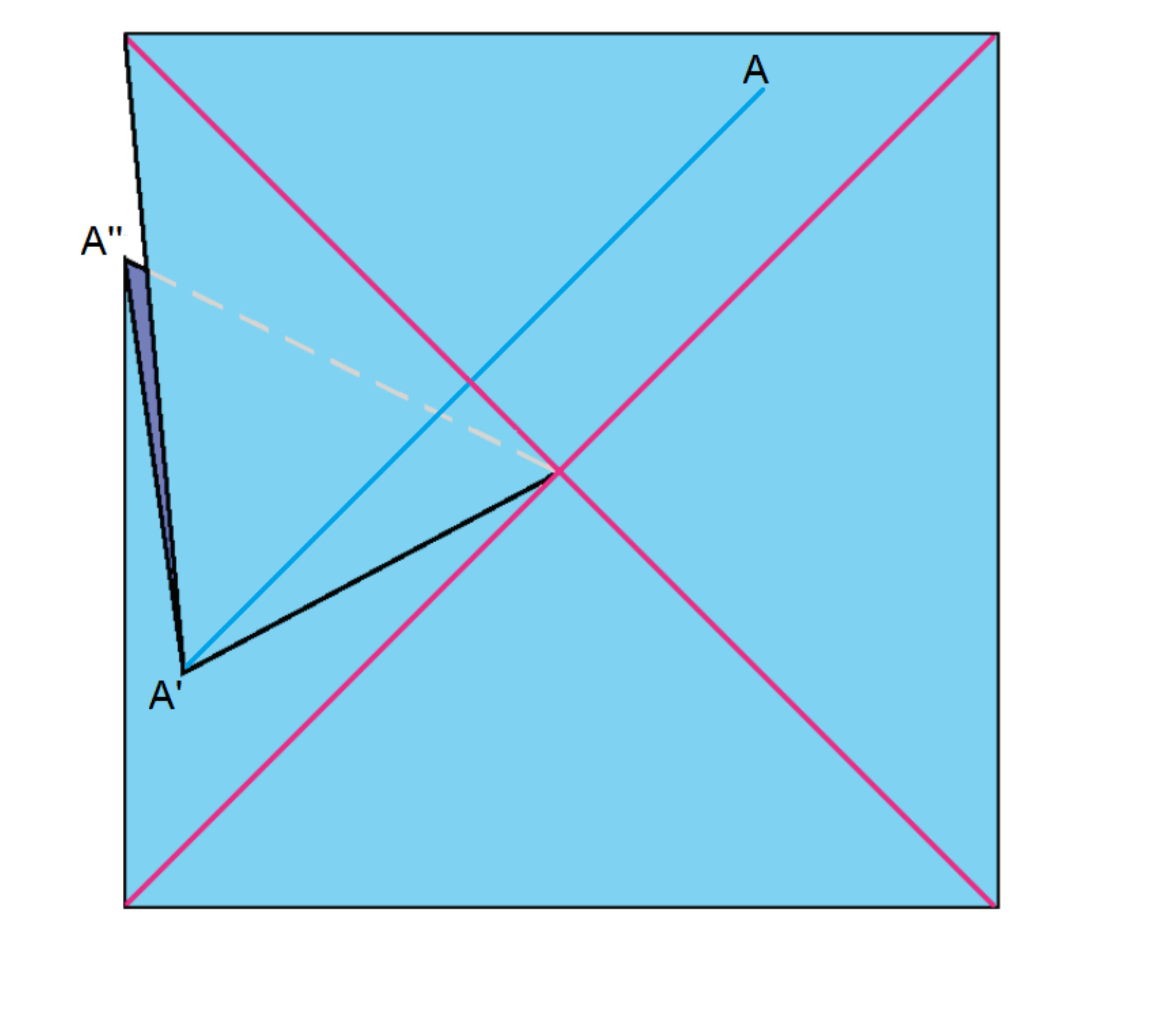}
\caption{ A timefold constructed from the left-side Hamiltonian. It is clear that a message can be sent to the region behind Bob's horizon. }
\label{halffold}
\end{center}
\end{figure}

One may also ask what the bulk geometry induced by acting with $A''$ looks like. The best answer is that the back-reaction induces a timefold \cite{Heemskerk:2012mn}. Since the left side has a Killing vector the timefold can be extrapolated into the bulk along constant time surfaces as in figure s \ref{halffold}.

From the figure one can see that a message can be sent to $A$ by acting with $A''.$

\bigskip

\subsection{A Paradox}

Precursors,  timefolds, and Einstein-Rosen bridges  lead to puzzling paradoxes that suggest a new form of uncertainty behind horizons. Consider the process illustrated in figure \ref{whoosh}. The quantum state of the two uncoupled CFT's  is a function of ``two-fingered" time, $t_L$ on the left side and $t_R$ on the right side:

$$|\Psi(t_L, t_R) \ra.$$

\begin{figure}[h!]
\begin{center}
\includegraphics[scale=.3]{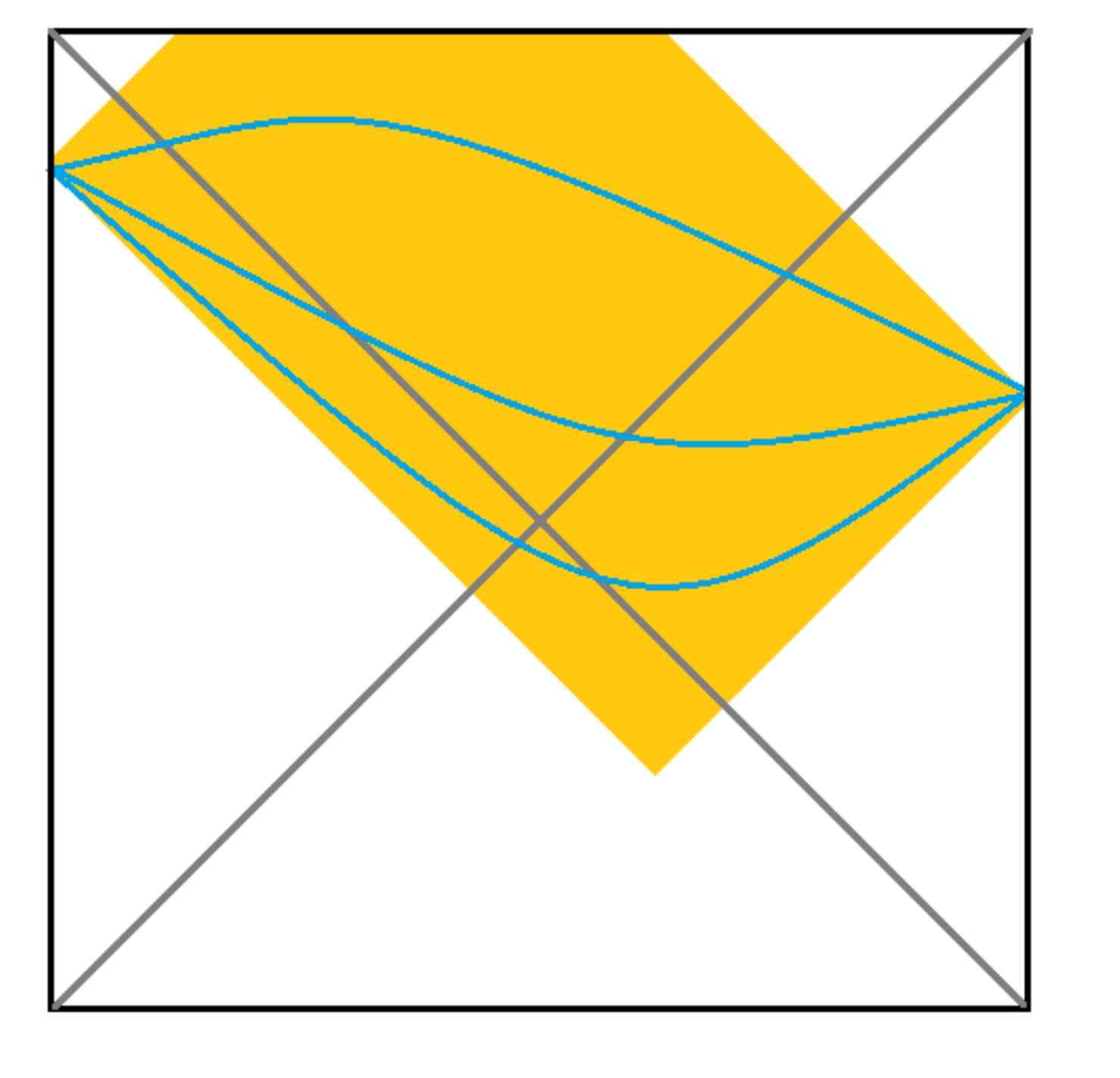}
\caption{A boundary wave function is dual to a Wheeler-DeWitt bulk history. The yellow region is foliated
by space-like surfaces that and at specific times on the boundaries. }
\label{wdw}
\end{center}
\end{figure}

\bigskip
\noindent
The gravity-dual of the wave function is a Wheeler-DeWitt  history for the bulk region containing  all space-like surfaces that end on the boundaries at $(t_L, t_R).$ Figure \ref{wdw} shows such a region.

\begin{figure}[h!]
\begin{center}
\includegraphics[scale=.3]{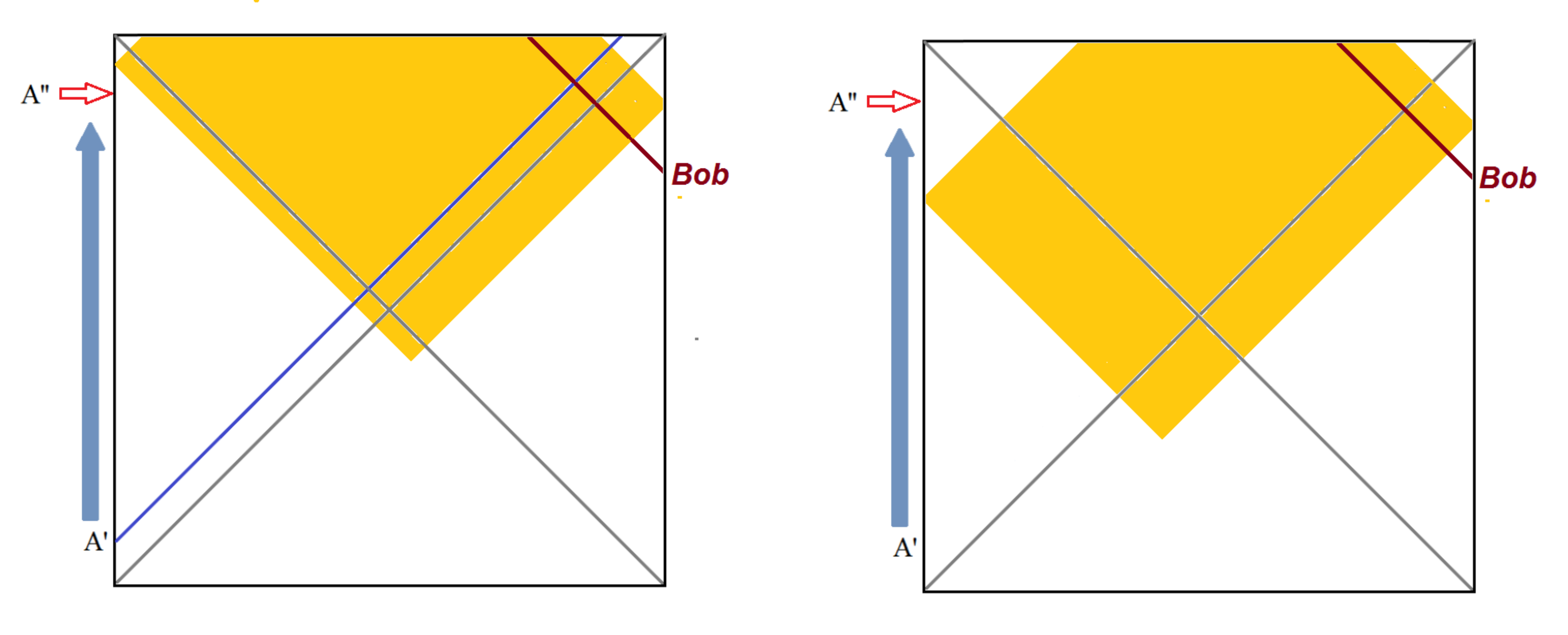}
\caption{Two WDW histories containing Bob. The left history  contains the measurement of $A''$ and the right history does not.}
\label{2wdws}
\end{center}
\end{figure}

Now let us reconsider the experiment in which Alice acts on the state with $A'',$ and Bob jumps into the right-side black hole in time to receive the particle from $A''.$  We can choose two WDW histories as in figures \ref{2wdws}. In both versions Bob's trip into the black hole is contained within the yellow region. In the first version the time on the left side is chosen before Alice measures $A''.$ In that case the state of the two CFT's is exactly the same as it  would have been had Alice  not measured $A''.$ Therefore the history should not contain a particle propagating from Alice to Bob.

On the other hand, in  the second version the CFT state on the left is defined after $A''$ has acted, and therefore should be exactly  as if $A'$ had acted. In this history Bob gets hit by a particle.

In the timefolded history in figure \ref{halffold} one sees that there is more than one Alice. On the first sheet (behind the others) Alice has not acted with $A'$ so she has every reason to believe that Bob will not detect a particle. On the third sheet (in front), Alice believes she has acted with $A'.$ Accordingly she believes Bob will get the message. This seems to be more Alice's problem than Bob's. The diagram seems to unambiguously say that Bob gets the message.

\begin{figure}[h!]
\begin{center}
\includegraphics[scale=.3]{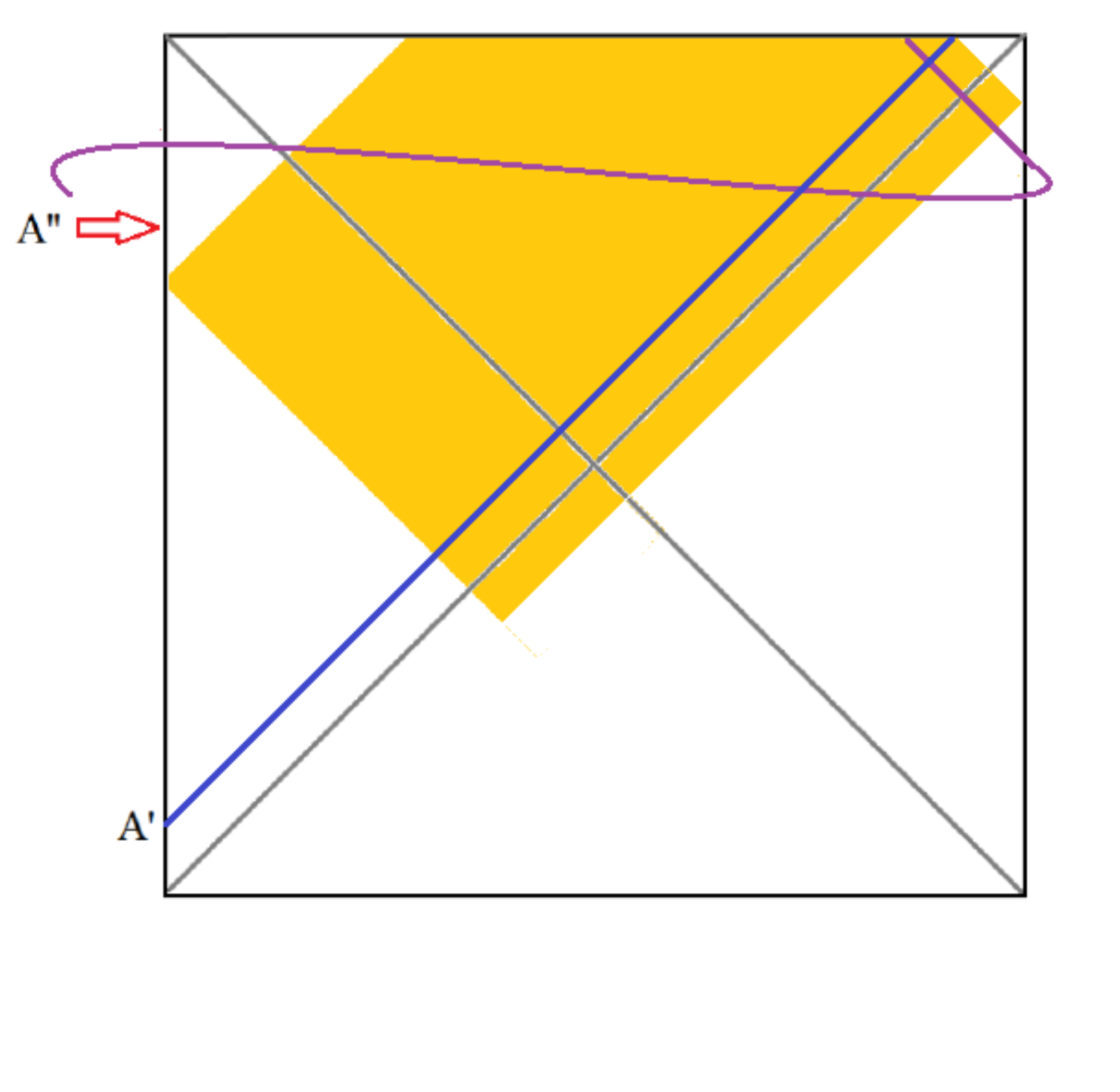}
\caption{Alice carries the result of her measurement to Bob's black hole and jumps in to check whether there is a particle at $A.$}
\label{Alice-flies1}
\end{center}
\end{figure}
The ambiguity goes away if Alice, after measuring $A'',$ takes the result to Bob just before he falls in, or better yet jumps in herself as in figure \ref{Alice-flies1}. In that case  the state of the system in either  case
contains information about the measurement.  Therefore  a particle should be found at $A$ as Alice expects.

What should we make of this ambiguity? Is it a contradiction or just a very unfamiliar new manifestation of quantum gravity? One possible answer is that $A''$ cannot be measured by Alice; that the powers needed to do the experiment are un-physical. For example, Harlow-Hayden computational complexity constraints \cite{Harlow:2013tf} could be invoked to  disallow measurement of $R_B$ before the black hole has evaporated. In the laboratory model there is no similar constraint  since the measurement of $A'$ can be done in polynomial time.

Another possible answer is that the ambiguity represents a serious contradiction which proves the firewall hypothesis.

The third answer is that behind-the-horizon physics is very unfamiliar and this is just a new kind of uncertainty  that we will get used to. Personally I like the last answer best.

\section{Gauge Invariance of $A$}
\subsection{A Gauge Theory Example}
\begin{figure}[h!]
\begin{center}
\includegraphics[scale=.3]{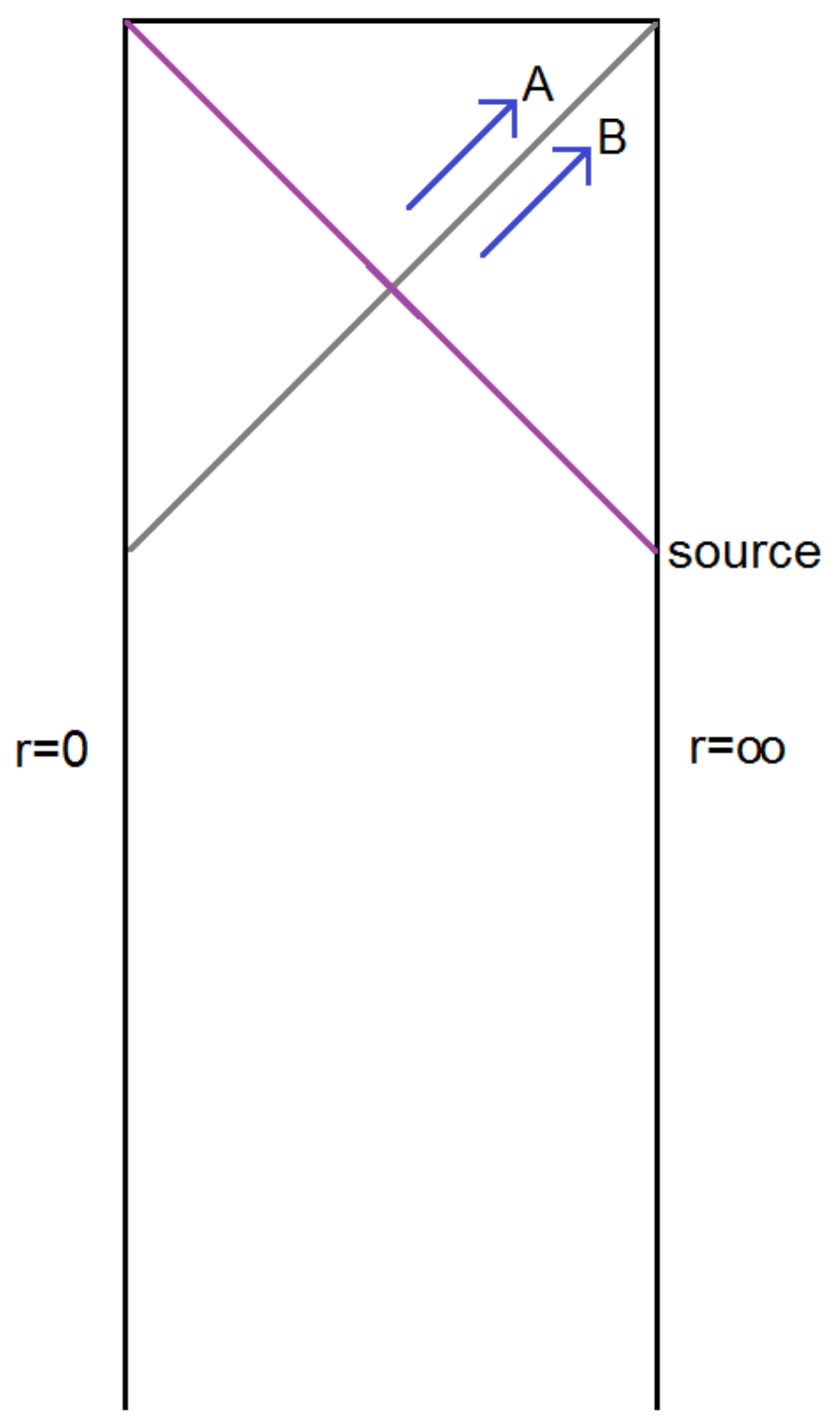}
\caption{Penrose diagram for a one-sided ADS black hole created by a shock wave from the boundary. }
\label{oneside}
\end{center}
\end{figure}

We would like to  know whether, and in what sense,  observations behind the horizon are encoded in the exterior description, by which I mean the S-matrix,  or in the case of ADS,  a dual  gauge theory.
Consider the case of a one-sided ADS black hole. For definiteness the black hole could be formed by activating sources on the boundary as in figure \ref{oneside}.

AMPSS argue that the modes $A$ behind the horizon cannot exist as operators in the CFT. The argument is based on a contradiction that was exposed in \cite{Almheiri:2013hfa}. The contradiction is between the canonical commutation relations

\be
[A^-, A^+]=1
\label{commutators AA}
\ee

\bigskip
\noindent
and the spectrum of states of a one-sided ADS black hole. A possible interpretation is that the interior geometry of a black hole does not exist. A somewhat weaker conclusion is that the black hole interior, although it exists, cannot be described by the degrees of freedom of the CFT. I will argue against both these conclusions.

The contradiction found by AMPSS will be explained after a digression about gauge theories.
In theories with gauge invariance there are generally fields whose properties conflict with the spectrum of physical states. I'll  review familiar example which has some similarities to our problem. Consider quantum electrodynamics on a spatial sphere of unit volume. The charged fields $\psi$ satisfy anti-commutation relations similar to \ref{commutators AA},

\be
\{\psi^-, \psi^+ \}=1
\label{commutators phiphi}
\ee

\bigskip
\noindent
This implies that the charge increases and decreases by one unit when $\psi$ acts. However, on a compact space the total charge must be zero. Thus the charged operators must annihilate every physical state, which of course is inconsistent with the commutation relations.

Returning to the AMPSS contradiction, the operators $A$ create and annihilate particles behind the horizon. If the black hole is older than the scrambling time it will be in equilibrium. The geometry will have a time-like Killing vector and a conserved Killing energy. Moreover, the Killing energy is negative for the behind-the-horizon quanta described by $A.$

This means that the creation operators $A^+$ \it decrease \rm the energy of a black hole, and as AMPSS noticed,  also decreases the entropy\footnote{See Section 4.3 for an explanation of the negative information carried by $A.$}; acting with $A^+$ must contract the number of states. In fact $A^+$ has to annihilate about half the states; a conclusion which is inconsistent with the commutation relations \ref{commutators AA}. This is similar to the electromagnetic case in which $\psi$ annihilates all states.

I believe there are two ways to deal with this paradox. One way is to accept the gauge dependence of the modes and construct a formal theory of negative information degrees of freedom. I will discuss this further in Section 4.3. The other way is to redefine the modes in a more operational way so that they are gauge invariant.

The electromagnetic puzzle can be resolved by adding a single un-physical degree of freedom corresponding to a uniform background charge density. Instead of defining the charge density to be $\frac{i}{2}[\psi^{\dag} \psi]$ we add an additional spatially constant component $Q$ where $Q$ is an integer valued operator. In addition we introduce unitary raising and lowering operators $e^{\pm i \theta}$ satisfying,

\bea
e^{ i \theta}Qe^{- i \theta} \eq Q+1 \cr \cr
e^{- i \theta}Qe^{i \theta} \eq Q-1
\label{theta}
\eea

\bigskip
\noindent
Then one defines the operators

\bea
\chi \eq \psi e^{i\theta} \cr \cr
\chi^{\ast} \eq \psi^{\ast} e^{-i\theta}
\eea

\bigskip
\noindent
which satisfy canonical anti-commutation relations

\be
\{\chi^-, \chi^+ \}=1
\label{commutators phiphi}
\ee

\bigskip
\noindent
From that point we quantize in the usual way except the the charge density has the extra term $Q.$

The operators $Q$ and $e^{i\theta}$ are not physical degrees of freedom. However they cancel out of any gauge invariant matrix element. In particular they cancel out of any amplitude between states in the zero charge sector. The lesson is not that the theory does not exist; it is that in order to represent the canonical commutation relations, one must add un-physical degrees of freedom that do not act in the physical Hilbert space.

Let us consider a related but different question in flat-space electrodynamics with bosonic charged fields. How does one measure a charged field  $\phi?$ The answer of course is that the phase of $\phi$ is not gauge invariant and therefore not observable. Measuring the phase requires an interaction that changes the conjugate variable; namely the charge. In other words measuring $\phi$ must involve interactions
that violate charge conservation.

However, it was understood long ago \cite{Aharonov:1967zza} that it is possible to measure the phase of $\phi$ relative to the phase of some charge reservoir. For example, the phase can be measured relative to the phase of a nearby superconductor. If the phase conjugate to the charge of the superconductor is $\theta$ then $\phi e^{i\theta }$ is an observable. Thus a measurement of $\phi$ requires an apparatus that includes such a reservoir.

What does this have to do with the modes $A?$
The relevant gauge symmetry is time translation symmetry. The conserved quantity is energy: the conjugate variable is time: and the analog of the charge reservoir is a clock. The need for a physical clock to define observables is familiar from the ``timeless" Wheeler DeWitt formalism.

The purpose of the clock is to activate Bob's measuring apparatus at the appropriate time to measure the mode $A$. Suppose Bob wants to measure $A$ at a given proper time after he jumps into the black hole from the boundary of ADS. He requires a clock to keep track of his time relative to clocks at infinity. The clock must have an energy uncertainty, and if we assume its energy is positive it must have a positive average energy. By the uncertainty principle, that energy must be of order the characteristic energy scale of $A$ which is equal to the inverse distance to the horizon. It is easy to show that this implies that the energy brought in by the clock is greater than the negative energy of the $A$ mode.
Thus, the observable gauge invariant operator, constructed out of $A$ and the clock, always has positive energy.

One way to see that the gauge invariant version of $A$ has positive energy is to write the contribution of $A^+$ to the observable field operator as

\be
A = e^{i\omega t} A^+.
\ee

\bigskip
\noindent
To make this gauge invariant we replace $ e^{i\omega t}$ by $ e^{i\omega t_c}$ where $t_c$ is the dynamical clock variable. The clock time $t_c$ does not commute with the Hamiltonian; in particular with the clock Hamiltonian, which is part of the full Hamiltonian. In fact $ e^{i\omega t_c}$ increases the energy by the same amount as $A^+$ decreases it.

There is another way to see that Bob needs a clock to measure $A.$ We could consider the case of Rindler space where Bob jumps into the vacuum, and attempts to measure a field-fluctuation averaged over a spacetime region of some size. For example, he might want to measure an electric-field vacuum fluctuation. He takes a tiny dipole with him; perhaps a molecule with an electric dipole moment in its ground state. When the molecule gets to the fluctuation it is excited by the electric field and Bob may read off information about the field from the effect on the dipole.

However this won't work. If the dipole is in its ground state it continually interacts with the vacuum fluctuations. The vacuum fluctuations are absorbed into the structure of the molecular ground state and into re-normalizing the parameters of the molecule. The resulting ground state is stationary and does not suddenly get excited by a vacuum fluctuation.
\begin{figure}[h!]
\begin{center}
\includegraphics[scale=.3]{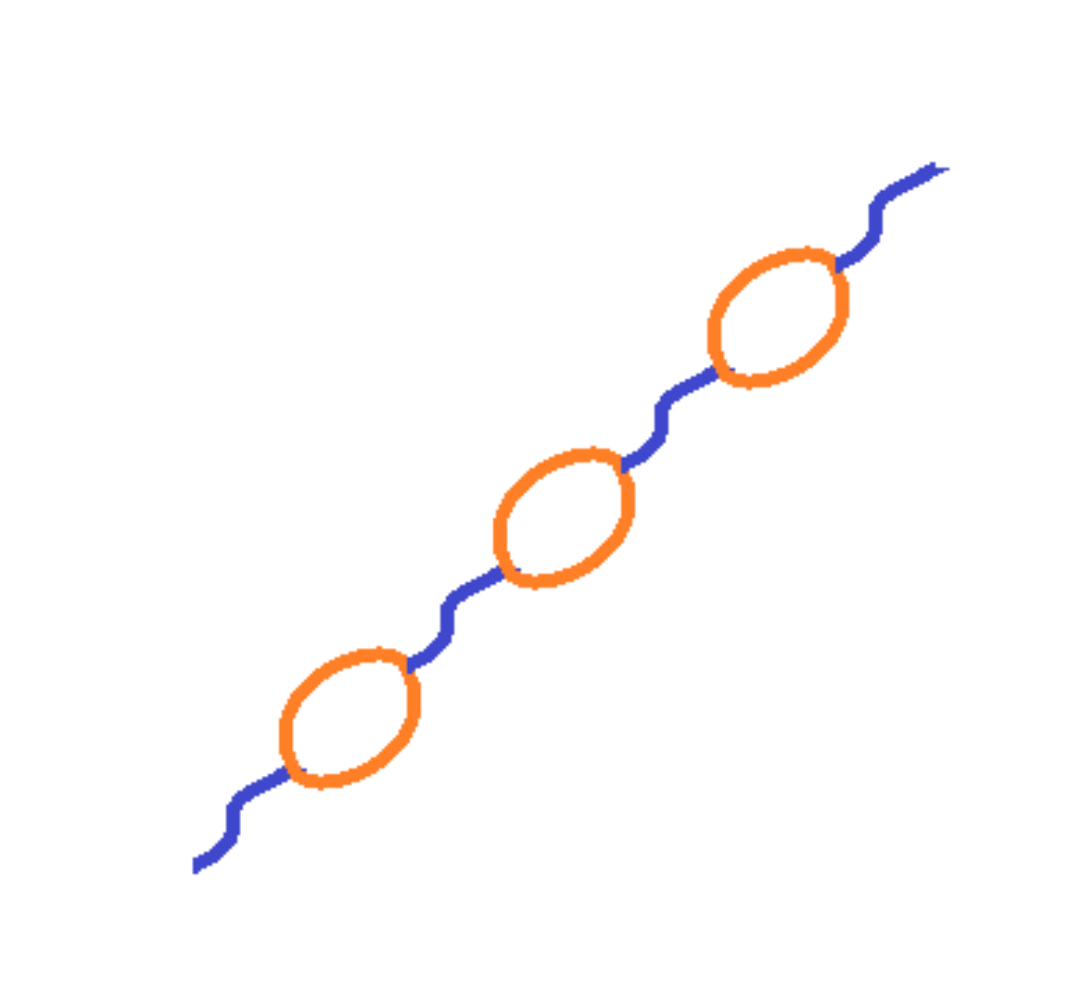}
\caption{A particle interacting with vacuum fluctuations. The vacuum fluctuations can be absorbed into the structure of the particle. }
\label{vacfluc}
\end{center}
\end{figure}
Bob needs to carry a clock with him, and at just the moment when he wants to measure the field, have his apparatus activate the dipole. The initial dipole moment may be set to zero, and at the appropriate time the apparatus separates the charges and activates the apparatus. This allows Bob to read off the electric field at the time and place of his choice.

\begin{figure}[h!]
\begin{center}
\includegraphics[scale=.3]{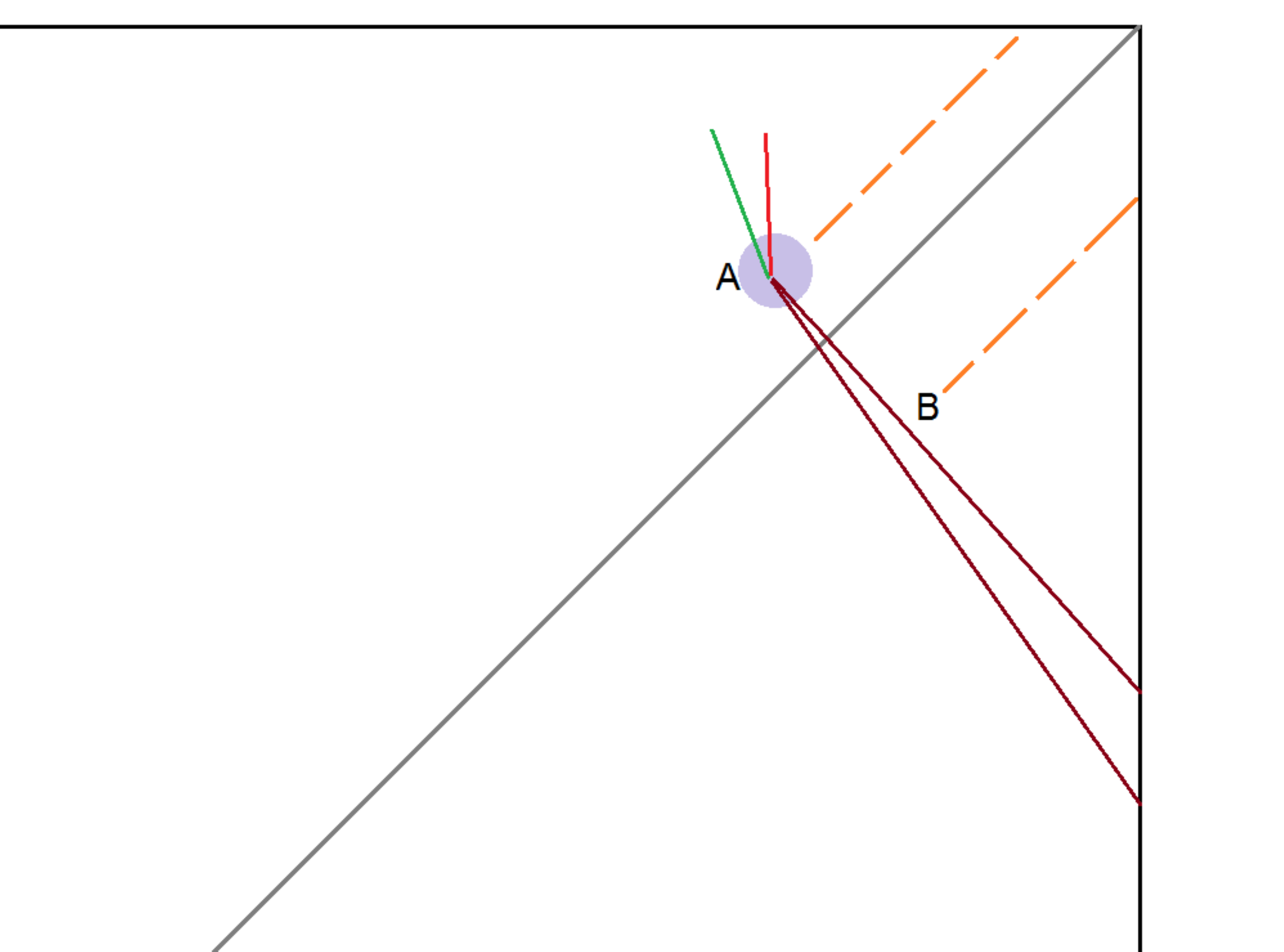}
\caption{The mode $A$ can be measured by colliding two particles at the site of $A.$ The incoming particles are neutral and the outgoing are an oppositely charged pair. By observing the pair one measures the vacuum fluctuation in the electromagnetic field at $A.$ }
\label{measure}
\end{center}
\end{figure}

Another way to accomplish the same thing is to shoot two neutral particle to the target location as in figure \ref{measure}. In this case the clock is the distance between the particles: when it goes to zero the apparatus is activated.
When they collide the particles produce a plus-minus pair of electric charges. The electric field acts on the pair, and from the final momenta of the pair Bob reads off information about the electric field. The magnetic field can be measured in a similar way and an appropriate combination of the two would allow a particular mode to be measured. But the actual operators that are being measured are a complex combination of the field and the clock degree of freedom.

Now we can ask about the encoding of $A$ modes in the boundary CFT. It should be obvious that only the gauge invariant versions of the $A$
are physical and can be encoded in the physical degrees of freedom of the CFT.

The argument also tells us that gauge invariant observables should be defined by reference to the operational procedures that actually measure them, and must include the apparatuses that do the measuring, such as clocks or particles. In fact the example given illustrates that the real observable associated with ${A}$ is not some field that has propagated up from the left through trans-Planckian modes, but rather an observable of a scattering process involving sub-Planckian particles, introduced from the right side. The $A$ modes welling up from trans-Planckian modes have no life of their own\footnote{Eva Silverstein has emphasized this same point. In her language the mode $A$ represents particles which are created by ``natural processes."}. What they represent when they interact with infalling matter is the radiative corrections to the systems that fall into the black hole within a scrambling time of $A.$

\begin{figure}[h!]
\begin{center}
\includegraphics[scale=.3]{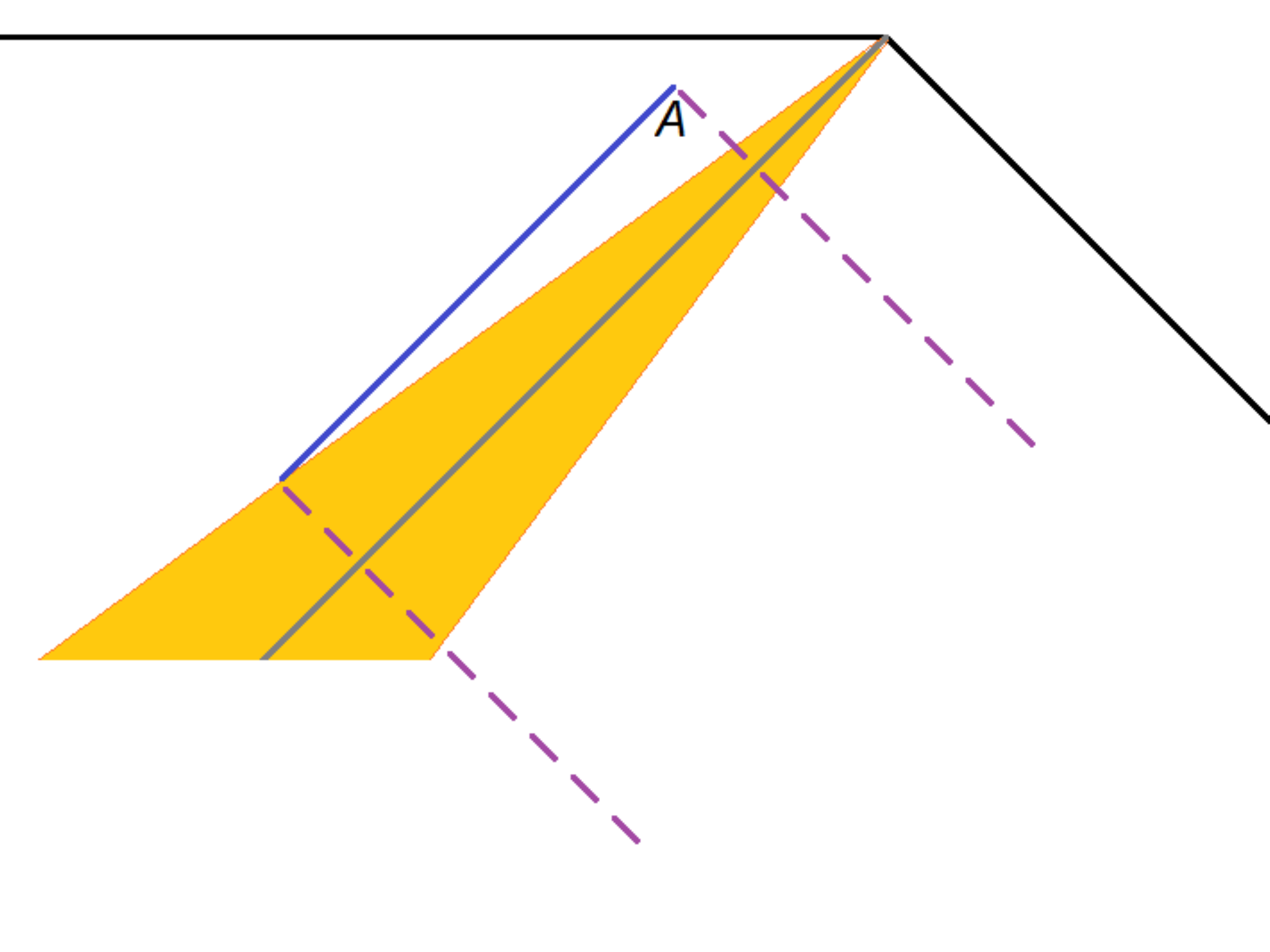}
\caption{The yellow region of the Penrose diagram shows the interior and exterior stretched horizon. The exterior stretched horizon is the usual place where the local temperature becomes Planck or string scale. The interior stretched horizon is its reflection behind the horizon.
The mode $A$ can be traced back until it enters the interior stretched horizon at which point it becomes Planckian. }
\label{stretch}
\end{center}
\end{figure}

Formally, the non-gauge-invariant operators $A$ can be run back along a path parallel to the horizon. At some point they will enter the interior stretched horizon as shown in figure \ref{stretch}.
The time that $A$ runs into the yellow region in figure \ref{stretch} is of order the scrambling time $t_\ast$ at which point it enters the trans-Planckian region. Going further back the mode will be squeezed into a distances much less than Planck-length from the horizon.

Let us ask whether we can make this backward extension of $A$  gauge invariant. It is obvious that a clock which can resolve the stretched horizon must measure times on shorter scales than the Planck time, and therefore carry energy greater than the Planck energy. In fact if we go backward from $A$ just two scrambling times, the clock must have  energy greater than the black hole mass. The back reaction would take us far out of the Hilbert space of the black hole, and there would be no sense in which the original mode was close to the horizon. Evidently $A$ cannot be run backward in a gauge invariant manner for more than the scrambling time scale. Therefore, in the one-sided case  no gauge invariant information can originate from trans-Planckian past history of the mode $A$.

The two-sided case of the laboratory model is different. In figure \ref{eternal} the backward extension of $A$ traces to $A'.$ $A'$ is a low energy operator on the left side. No additional clocks, other than those at the boundary, are needed to define it gauge invariantly.

The upshot is that AMPSS is correct \cite{Almheiri:2013hfa}\cite{Joe} that the naive $A$ modes  carry negative energy and cannot be realized in the physical Hilbert space of the one-sided CFT. Moreover gauge invariant versions of them cannot be run back for more than a scrambling time. But this does not imply that there is no black hole interior. It implies that the gauge invariant degrees of freedom behind the horizon represent low energy matter that came across the horizon less than a scrambling time earlier. The effects of the $A,B$ modes on the infalling matter is equivalent to radiative corrections.

\subsection{Young Black Holes}

The arguments give above were for relatively old one-sided black holes. The situation for young black holes is somewhat different.

\begin{figure}[h!]
\begin{center}
\includegraphics[scale=.3]{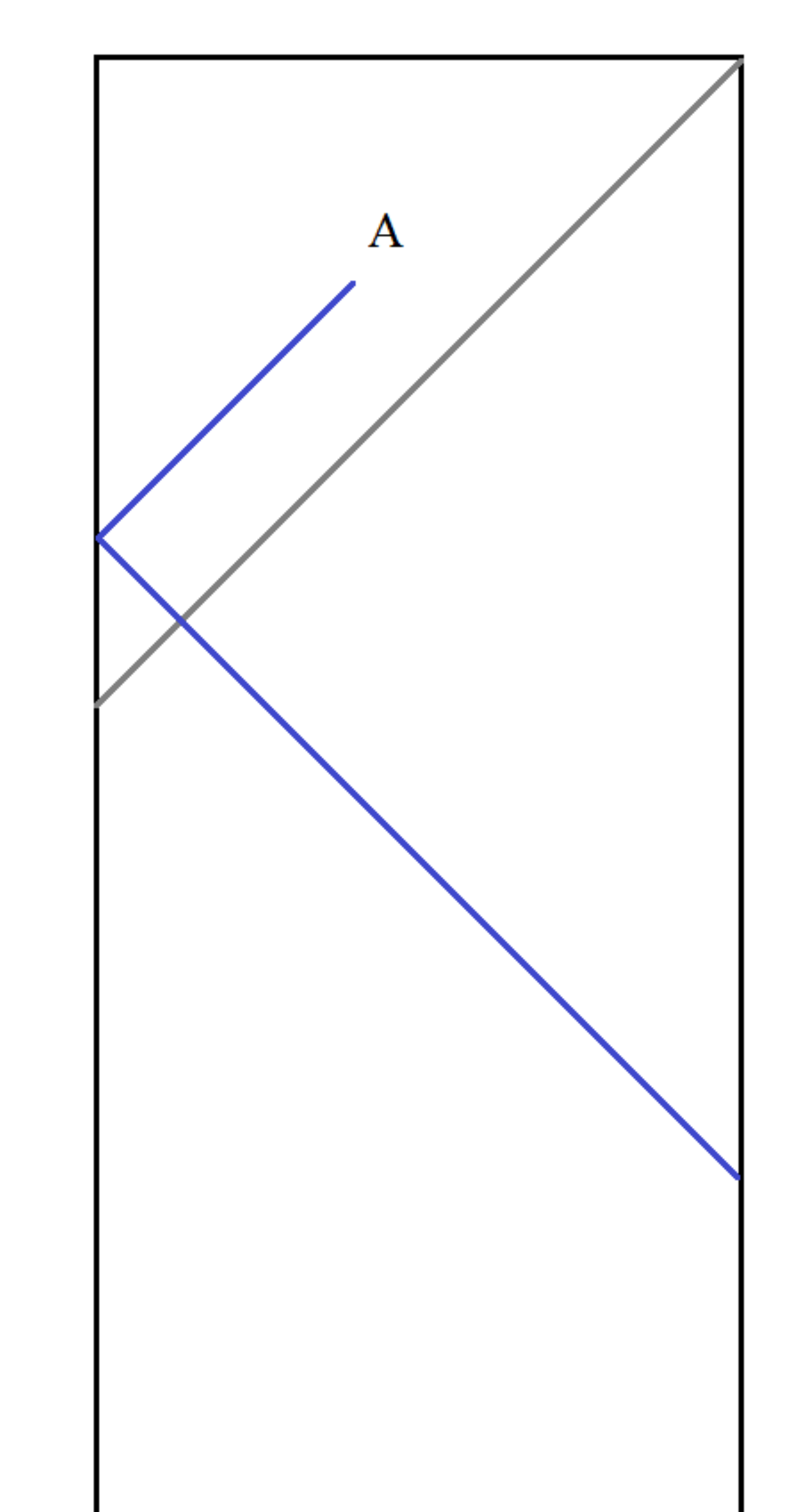}
\caption{ In the case of a young black hole the mode $A$ can be defined gauge invariantly by ray-tracing it back to the boundary. There is no real distinction between modes propagating up from the lower-left and things thrown in from the right.}
\label{Young}
\end{center}
\end{figure}

Consider a one-sided black hole younger than the scrambling time. In that case the $A$ modes may be defined in a gauge invariant manner. In figure \ref{Young}
the mode $A$ is defined by running it back along the light-like trajectory to the boundary or to past light-like infinity in the asymptotically flat case.

First of all the AMPSS argument about negative energy does not apply before the scrambling time. During the young phase the black hole is not in equilibrium and there is no approximate time-like killing vector. The quanta of $A$ do not carry negative energy
so there is no clash between the commutation relations of $A$ and the energy spectrum.
The clash only occurs after the scrambling when the black hole has settled down to its equilibrium state.
Note that in the young case there is no clear difference between entering the black hole from the right and propagating up from the far left. More on this in Section 4.3.

\subsection{Does the CFT Know What Lies Behind the Horizon?}
Polchinski and Bousso have both raised the question: Does the dual CFT (Or the S-matrix in the flat-space case) contain the information as to the fate of a system behind the horizon? They formulate it this way: Suppose Heisenberg's cat ( in his sealed box) falls through the horizon of a black hole. Is there an operator in the CFT dual description whose measurement would reveal whether the cat is still alive two seconds after falling through? The best way to answer this question is to consider a specific map between the interior and the exterior.

\begin{figure}[h!]
\begin{center}
\includegraphics[scale=.3]{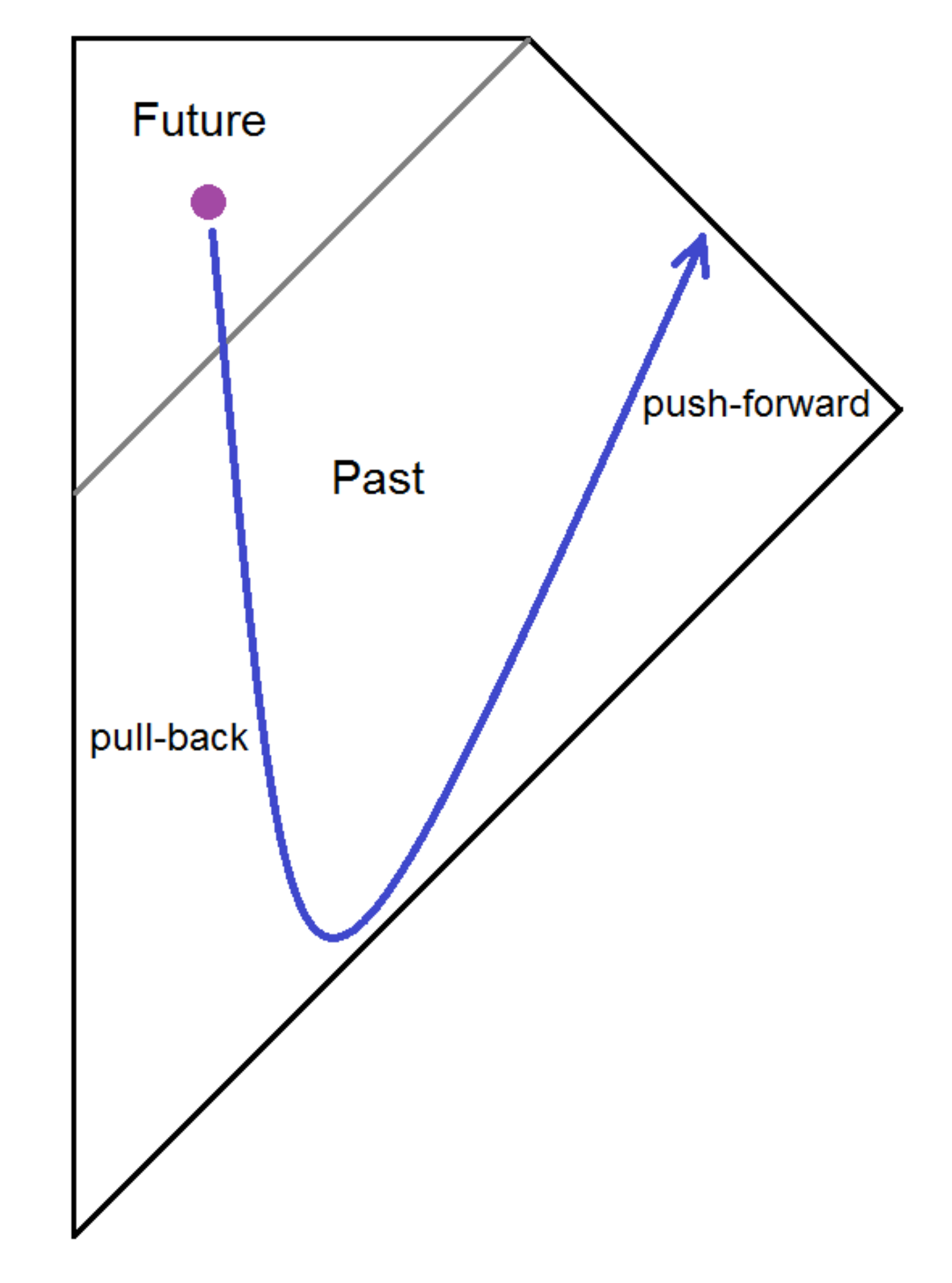}
\caption{ The region behind the horizon is to the future of the exterior. The purple dot indicates an observable behind the horizon. By using the low energy equations of motion in the infalling frame the observable can be expressed in terms of operators at past light-like and time-like infinity. This is the pull-back half of the construction. The push-forward is accomplished by conjugating with the S-matrix. }
\label{pullpush2}
\end{center}
\end{figure}

The
pull-back---push-forward map \cite{Freivogel:2006xu} may or may not be useful for old black holes, but it is straightforward for the observables of a very young black hole (younger than the scrambling time). This should be good enough to answer the question posed above.
The map is based on the idea that the region behind the horizon is in the future of the exterior of the black hole. Figure \ref{pullpush2} illustrates the sense in which this is true for the flat space case.

 The pull-back procedure takes a low energy observable behind the horizon and solves for it in terms of earlier exterior degrees of freedom in the remote past. It is then pushed forward using the S-matrix, or in case of gauge-gravity duality by using the  gauge theory equations of motion.
The pull-back---push-forward is a special case of a timefold in which half the fold takes place in the infalling frame and half in the exterior frame.

Consider an experiment behind the horizon such as measuring $X$ by means of a scattering operation in which particles come in from the outside and interact at $X.$ Is there an operator in the Hawking radiation (or near the boundary of ADS) which can be measured and will tell us the result of the experiment behind the horizon? Ordinary causality tells us that the answer is no. But that's not the end of the story. The pull-back---push-forward allows us to relate the interior observables to exterior operators in the Hawking radiation.

We begin the mapping procedure by running the observable $X$
back to past infinity.

\be
\tilde{X}_{in} = U^{\dag} X U
\ee

\bigskip
\noindent
where $U $ is the low energy evolution operator that runs  back to past infinity in the bulk theory. The tilde indicates that the operator is defined as either an operator on past light-like infinity or as an operator in the dual CFT in ADS/CFT.

Next run the operator forward using the $S$ matrix to obtain $
\tilde{X}_{out}$

\bea
\tilde{X}_{out} \eq S^{\dag} X_{in} S \cr \cr
\eq (S^{\dag} U) X (U^{\dag}S)
\eea

\bigskip
\noindent
In the case of ADS we could run the operator up to time zero if we preferred and define an operator

\bea
\tilde{X}(0) \eq V^{\dag} X_{in} V \cr \cr
\eq (V^{\dag} U) X (U^{\dag}V)
\eea

\bigskip
\noindent
where $V$ is an evolution operator in the dual CFT in the gauge/gravity dual case.
The question is in what sense $\tilde{X}$ has the same information as $X?$ The answer is that the probabilities for different outcomes are the same for both. To say it differently, the expectation values of all functions of $X$ are the same as those of $\tilde{X}_{out}$. But does measuring $\tilde{X}$ actually tell us what happened behind the horizon?

The answer is no as can seen from the example of the free particle in Section 4.1. Consider the probability to observe a particular value of $x$ if it is measured at time $t=0.$ The answer is exactly the same as the probability to measure the $x-pt$ at time $-t.$ But it is not right to say that measuring $x-pt$ at time $-t$ tells you what the result of measuring $x$ at time $t.$ In fact a measurement at $-t$ would disturb the system and interfere with the experiment at time $0$. The only correct statement is that the probability to get a given value $x_0$ if you measure $x$ at time $0$ is the same as the probability to get $x_0$ if you measure $x-pt$ at time $-t.$ In other words the mapping of operators from one time to another preserves the statistics.

The same is true for measuring $X$ or $\tilde{X}.$ For example a measurement of $\tilde{X}(0)$ has the same probability distribution as a measurement of $X$ behind the horizon. But the measurement of $\tilde{X}(0)$ disturbs the interior measurement of $X.$

It is best to not think of the region behind the horizon as literally part of the CFT.
 Instead, the CFT is a machine for constructing initial conditions on the light like horizon from CFT data, which can  be continued behind the horizon. In this way of thinking the region behind the horizon is to the future of everything directly described by the CFT\footnote{This point was emphasized to me on several occasions by Juan Maldacena.}.

 To put it another way, the spacetime dual to the CFT is not geodesically complete. The horizon is the bulk representation of gauge-theory time $t= \infty;$ and the firewall question is about the consistency of completing the spacetime using bulk equations of motion.

 One last point about pull-back---push-forward: it has been argued that it cannot make sense for black holes older than the scrambling time. The reason involves the pull-back through the yellow region in figure \ref{stretch}; the running of $A$ all the way back involves ultra-trans-Planckian  physics. But if there is no gauge invariant information in the trans-Plankian extension of $A,$ then the only gauge invariant information is what entered from the right side within a scrambling time of $A.$

 The exception to this is information that entered through an Einstein-Rosen bridge, and which can be traced back to low energy degrees of freedom at the other and of the bridge. The pull-back may have to include extrapolation through the bridge in the case that the black hole is maximally entangled with a distant system.

\sc
\section{{Black Holes as Messages}}
\subsection{Easy and Hard Operators}
The AMPSS commutator argument is another paradox that may have a lot to teach us \cite{Almheiri:2013hfa}. In particular it leads to a very interesting perspective on the meaning of the black hole interior; namely, \it the interior as a fault-tolerant quantum message.\rm \ The original purpose of the argument was to refute  $A=R_B$ which in any case cannot be literally correct. But AMPSS leads to another puzzle about the difference between ``easy" and ``hard" operators.

Let $E$ represent an operator acting on a single localized radiation mode\footnote{We model the modes by qubits. In AMPSS \cite{Almheiri:2013hfa} the notation for such modes is $e.$} of the Hawking radiation.
$E$ can stand for a number of things, but here we take it to mean ``easy "  as in \it easy to measure. \rm By contrast $R_B$ is a ``hard\footnote{The term hard in some contexts means high energy as in hard X-rays. In this paper hard refers to the difficulty of measurement and not the energy.}" operator, thoroughly de-localized and composed of all the early radiation modes. It is quite possible that $R_B$ is so hard to measure that it would take an exponentially long time \cite{Harlow:2013tf}, but for now that complication will be ignored.

Given that the black hole and its radiation are in a scrambled state, AMPSS show that \it any \rm $E$-qubit fails to commute with \it every \rm $R_B$-qubit, and that the commutators are of order $1.$ Thus, according to AMPSS, measuring or acting with any $E$ will disturb $R_B,$ and therefore send a particle to $A.$ Moreover disturbing \it any \rm $E$ will corrupt \it every \rm $A$ and therefore create a firewall. Measuring an $E$ can be as simple as having a single Hawking photon scatter off a dust particle.

The counterargument \cite{Verlinde:2012cy}\cite{Papadodimas:2013wnh} is that acting with $E$ only changes the identification of the system that purifies $B.$ No local operation performed on a subsystem can change its entanglement entropy with the complementary subsystem.
Thus if $E$ is measured then the definition of $R_B$ becomes modified by the interaction. The new $R_B$ includes the degrees of freedom that became entangled with the radiation when $E$ was measured. According to this view measuring $E$ only ``re-adjusts the code," but does not send messages to the interior of the black hole.

On the other hand measuring $R_B$ is somehow different since it would send a signal to the interior of the black hole (subject to the ambiguities of Section 3). This is unsettling because $R_B$ is made of easy operators, albeit a great many of them. One can resolve the problem of easy operators by saying that they alter the code; and one can solve the problem of hard operators by saying they create particles behind Bob's horizon. The hard thing  is to resolve both in a consistent way.
It is critical for unraveling  the AMPSS commutator paradox to identify a clear mathematical criterion that distinguishes
hard and easy operators. The  following sections are an attempt to do so. It is obviously incomplete.

\bigskip

\bigskip

\subsection{ Messages, Reference Systems, and Codes}

\bf NOTE ADDED TO V.2 

 After posting the original version of this article I have begun to doubt that  error correctability  is the right way to distinguish easy and hard. The conjecture on page 41 gives a sufficient condition for an operator to be easy but it does not give a sharp criterion for when it is hard.
I  don't think error correctability is the essence of the problem.

I described a more interesting set of criteria in  \cite{Susskind:2013aaa}. According to this paper the distinction is more a matter of chaos and complexity than correctability. The implications of \cite{Susskind:2013aaa} are that hard operators are a lot harder than error (un)correctability would imply.

What follows is the original version based on error correction. I have left it in place because I think most  of the ideas may be of some value.

\rm

In \cite{Maldacena:2013xja} (see also \cite{Verlinde:2012cy})
it was speculated that the difference between hard and easy operators is similar to the difference between correctable and un-correctable errors in the encoding of a message\footnote{It is not necessary that there be someone to do the error correcting. I am using error correctability as a mathematical property which may be used to distinguish hard and easy.}.  I will expand on this
idea.

Let's begin with some general concepts about encoding quantum messages. I will make use of four systems of qubits, $\a,$ $\n,$ $\b,$ and $\c.$

\bi

\item The system $\a$ contains $k$ qubits and it represents the original message before being encoded in a bigger system.

   \item $\n$ is a much bigger system containing $N>>k$ qubits. Initially  it is in some fixed \it initializer \rm  state ${|\n\ra}_0.$
The initializer state is independent of the message. For definiteness we can take it to be

\be
{|\n\ra}_0 = |00000000....\ra,
\label{initialize}
\ee

\bigskip
\noindent
but any state will do.

 \item $\a$ and $\n$ will be tensored together  to form a $(k+N)$-qubit \it code system \rm  $\c.$ The code system will be scrambled to make the code fault-tolerant.

     \item $\b$ is isomorphic to $\a$ and also consists of $k$ qubits. It is called the reference system and will be defined shortly.

\ei

What exactly  is meant by a quantum message? A $k$ qubit message is just a state of $k$ qubits. In other words it  is a vector in the space of states of the system $\a,$

$$|m\ra.$$

\bigskip
\noindent
To encode the state in a fault tolerant manner we combine $\a$ with  $\n$ to form the code $\c.$ The initial state of the code is a product state:

\be
|m\ra \otimes {|\n\ra}_0.
\ee

\bigskip
\noindent
The message at this stage is no more fault-tolerant than it was to begin with; the loss of any qubit of $\a$ is not recoverable.
The message can be made fault-tolerant by scrambling the code system. This is done by applying a ``scrambler---a random unitary operator $U_{scr} $---to the entire code system $\a \otimes \n.$

\be
|m\ra \otimes {|\n\ra}_0 \ \to \ U_{scr} |m\ra \otimes {|\n\ra}_0
\ee

\bigskip
\noindent
The original message is now embedded in the code in a manner extremely difficult to extract, but also very fault tolerant. One can accidentally flip or even lose a large number of qubits and still recover the message with high fidelity.

Another concept of importance is the reference system $\b$. The reference system enables us to go beyond encoding a particular message. It allows us to encode all possible $k$-qubit messages simultaneously.  Before scrambling, each qubit $A$ is partnered with a reference qubit $B$ to form a Bell pair, $$|00\ra + |11\ra.$$ The qubit $A$ may be called the purification of $B,$ meaning that the $A,B$ subsystem is in a pure state.

The initial state of the entire system is

\be
|\Psi_0\ra = {|\n\ra}_0 \otimes (|00\ra + |11\ra )^{\otimes k}
\label{initial}
\ee

\bigskip
\noindent
Any particular message can be recovered by projecting onto a state of the reference system.

Once the initial state \ref{initial} has been set up, we can make the encoding fault-tolerant by again scrambling the code system $\c$ with the scrambler $U_{scr}.$

\bea
|\Psi\ra &=& U_{scr}|\Psi_0\ra \cr \cr
\eq U_{scr} \ \left\{ {|\n\ra}_0 \otimes (|00\ra + |11\ra )^{\otimes k} \right\}
\label{scramstate}
\eea

\begin{figure}[h!]
\begin{center}
\includegraphics[scale=.5]{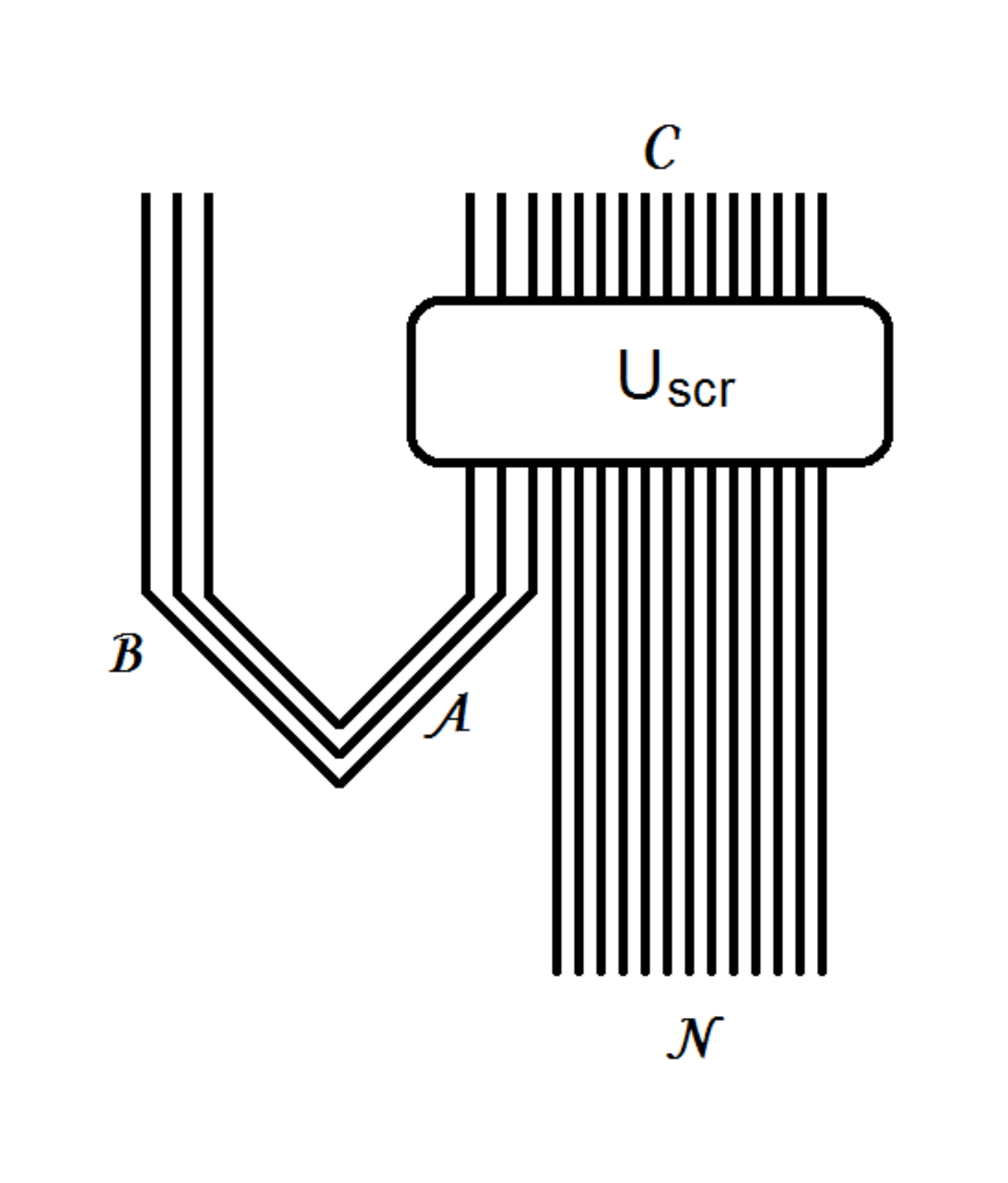}
\caption{ Encoding a message in $\c.$ The set of messages $\a$ are entangled with the reference system $\b.$ Then $\a$ is combined with a much larger system $\n$ and scrambled by $U_{scr}.$}
\label{circuit}
\end{center}
\end{figure}

\bigskip
\noindent
The scrambler $U_{scr}$ acts only on the $\a,\n$ system. On $\b$ it acts as the identity.
The entire encoding operation is indicated in the circuit diagram \ref{circuit}

What makes the code fault-tolerant is that one can erase rather large number of qubits and still preserve the almost exact maximal entanglement between $\b$ and the code. In other words a large number of qubits can be traced over, and there will still be a purification of any $B$-qubit in the remaining code subsystem.

The method of encoding that I described is conventional: a message of size $k$ is scrambled into a system of size $N$ to produce a code of size $(N+k).$ However this is not what is required to describe the evolution of  black holes. What we need is a  procedure which begins with a $k$-qubit and a system of $N$ qubits to  produce a code comprised of $(N-k)$ qubits. It is as though the message were counted as having $-k$ qubits. In that way, when the $k$ reference qubits are added, the total qubit count does not increase.

An increasing qubit count makes sense as a very young black hole undergoes its first episode of scrambling. During that phase new degrees of freedom are being liberated and the entropy increases from the relatively small value of the infalling matter to the much larger Bekenstein Hawking entropy. But for a black hole that has already scrambled, the qubit count is basically constant if we include both the  black hole and its Hawking radiation\footnote{There is a small increase of entropy as the black hole radiates \cite{Page:2013dx}. It can be taken into account but I will ignore it here.}. To model this we will have to formally count the message qubits as carrying negative information after the scrambling time.

\subsubsection{Negative Bits}
In order to describe black holes after the scrambling time we will need a modified encoding procedure in which the $k$-qubit message is encoded in a code of size $N-k$ instead of $N+k.$ For $N>>k$ there is plenty of available code space but it's not obvious how to accomplish the goal. In particular, a mapping from the Hilbert space of  $N+k$ qubits to a state of $N-k$ qubits cannot be done by a unitary mapping. If the mapping is not unitary there is no obvious reason why the message qubits will be encoded faithfully in the final code.

\begin{figure}[h!]
\begin{center}
\includegraphics[scale=.5]{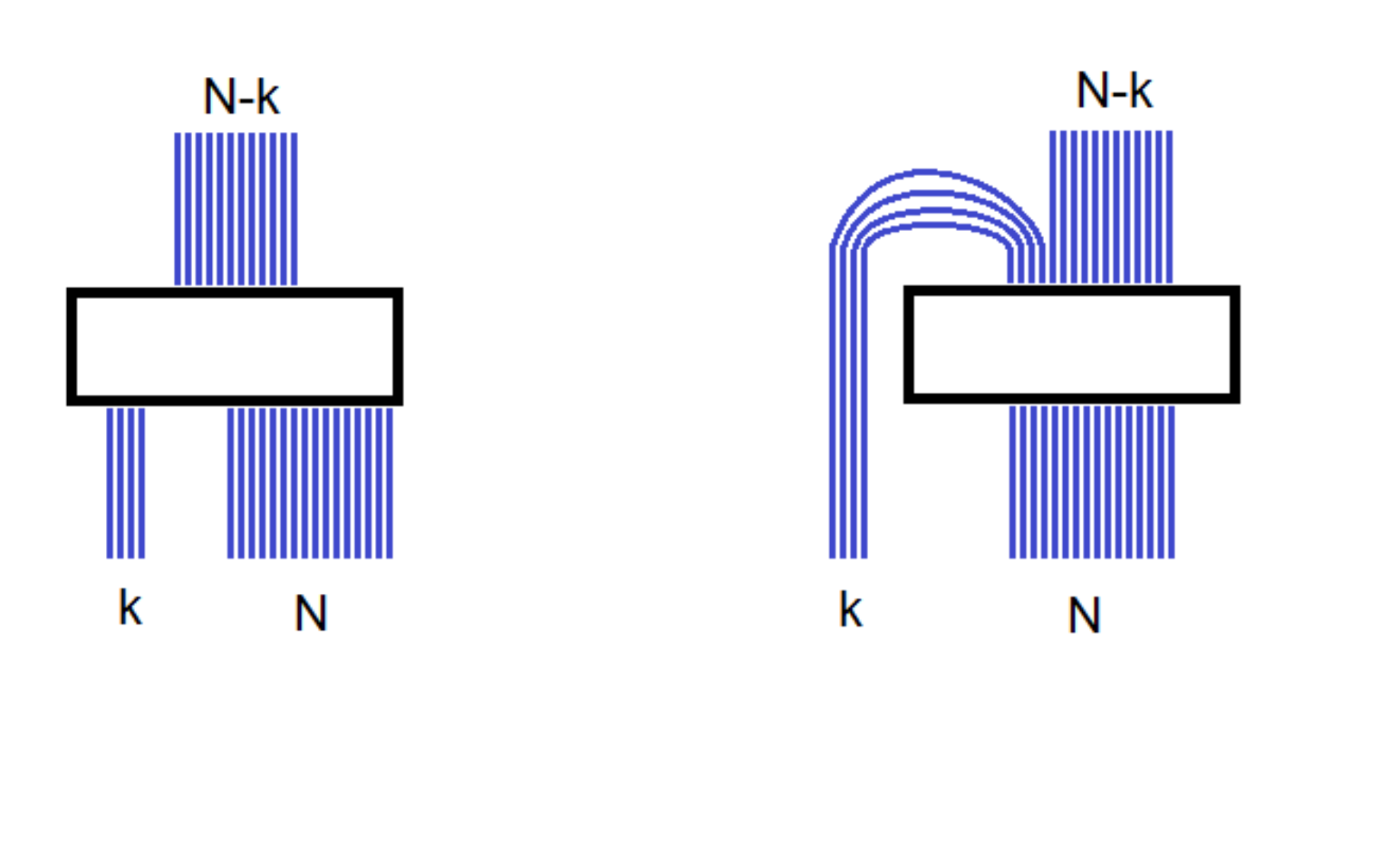}
\caption{ Defining a $2^{N+k} \times 2^{N-k}$ matrix in terms of a unitary $2^{N} \times 2^{N}$ matrix by crossing legs from the initial to the final state.}
\label{rectangular}
\end{center}
\end{figure}

The procedure can be accomplished using  a rectangular
$$2^{N+k} \times 2^{N-k}$$ scrambling matrix, defined in terms of a related square $$2^{N} \times 2^{N}$$ matrix. To do this we just transfer $k$ qubits from the initial to the final state according to the scheme shown in figure \ref{rectangular}.

To illustrate for the case $k=2$  we take a $2^{N} \times 2^{N}$  square matrix and rewrite it as a rectangular $2^{N+k} \times 2^{N-k}$ matrix. For the case of $N=4$ and $k=2$ it would look like this:

\be
U^{i_1 i_2}_{j_1 j_2 j_3 j_4 j_5 j_6} \ = \ \sum_{i_3 i_4}  U^{i_1 i_2 i_3 i_4}_{j_1 j_2 j_3 j_4 } {\cal{T}}_{i_3 j_5}
{\cal{T}}_{i_4 j_6}
\label{shift}
\ee

\bigskip
\noindent
where ${\cal{T}}$ is the anti-unitary operator which reverses the sign of all Pauli matrices.
The first $k$ indecies of the initial state have been shifted to the final state to make a rectangular $$2^{N+k} \times 2^{N-k}$$ matrix. This is  another example of a timefold.

\begin{figure}[h!]
\begin{center}
\includegraphics[scale=.5]{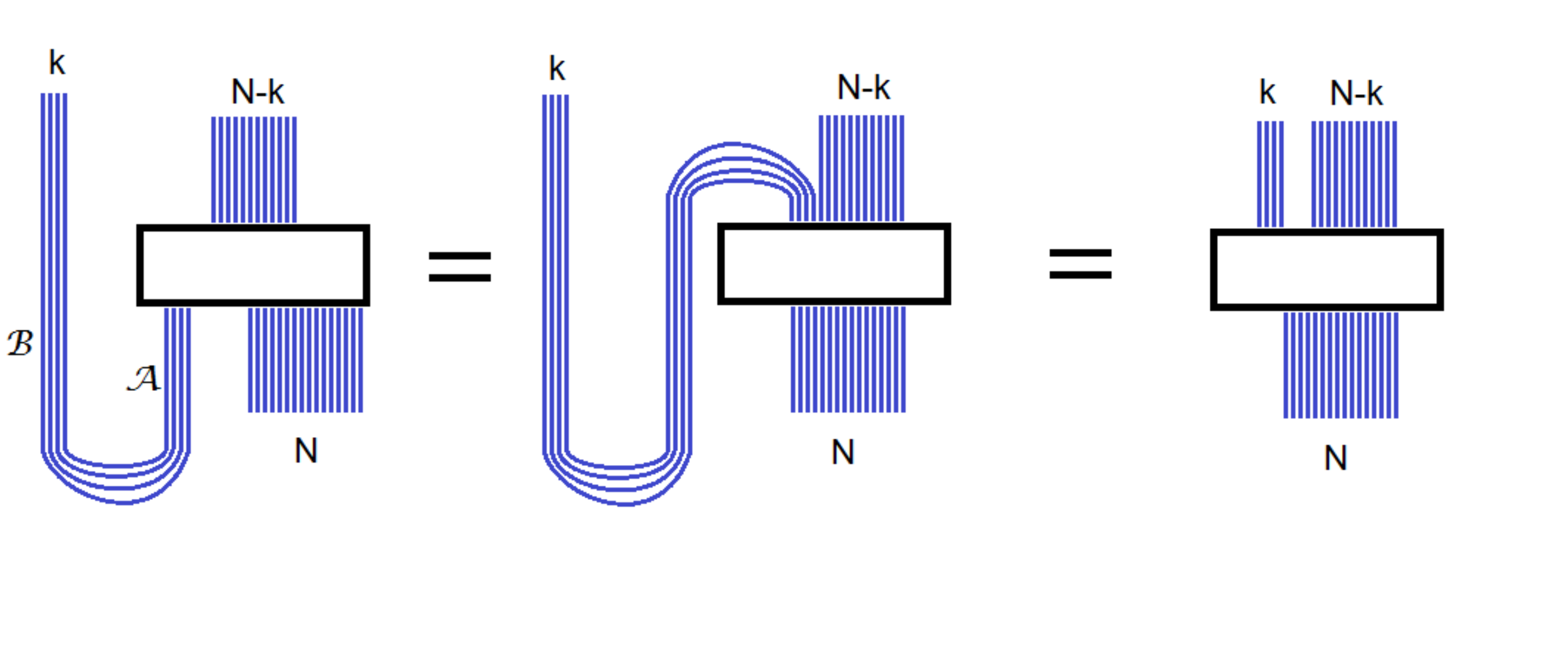}
\caption{ A mapping from $N$ qubits to a system of $k$ reference qubits combined with $N-k$ code qubits.}
\label{toomuch}
\end{center}
\end{figure}

Once we have the $$2^{N+k} \times 2^{N-k}$$ matrix we can use it along with the reference system to construct a mapping from $N$ qubits to a system of $k$ reference qubits combined with $N-k$ code qubits. This is shown in figure \ref{toomuch}

The procedure allows us to formally count message qubits as if they were negative. When the $A,B$ entangled pairs are created, they count as a total of zero qubits. Scrambling the message  with the $N$ initial qubits results in  a code of $N-k$ qubits and $k$ reference qubits. The reference qubits are purified by the code so that the message is encoded faithfully and fault-tolerantly. The formal trick of counting the message qubits as negative can be compared with the discussion in Section 3.

How do we know that the $k$-qubit message is encoded faithfully in the $(N-k)$-qubit code, given that the non-square matrix is not a unitary mapping? The answer is that if the $2^{N} \times 2^{N}$ unitary matrix is a random scrambler, then by Page's argument \cite{Page:1993df} one can be sure  that at the end,  the $k$ qubits of the reference system are almost maximally entangled with the code. There is however a difference between the encoding procedure  in equations \ref{initial}\ref{scramstate} where the $A$-qubits are counted as positive. In that case, by construction, the reference qubits are exactly maximally entangled with the code. The encoding defined by figure \ref{toomuch}, in which the $A$-modes are counted negatively, gives almost maximal entanglement, accurate to  exponentially small errors.

In the first type of traditional encoding, the faithfulness (as opposed to the fault-tolerance) of the code does not require scrambling. It only requires unitarity. In the second type of encoding where the code size is $(N-k)$ scrambling is essential to both the approximate faithfulness of the code and the fault-tolerance.

\subsection{Correspondence with Black Hole}
Let's these ideas to an old black hole; one which has evaporated far past the Page time, but which is still large enough to be classical.
The correspondence is as follows:
\bi

\item The zone of the black hole is the reference system $\b.$ Its entropy is $k.$

\item The Hawking radiation together with  the stretched horizon comprise the code  system, consisting of $N-k$ qubits.

\item The interior of the black hole is not represented by an independent set of degrees of freedom. It is represented by the message $\a$, which after scrambling is mixed into the code. A scrambled message of this type is fault-tolerant for a wide class of errors, but is also extremely difficult to decode.
\ei

The initializer state of ${|\n\ra}_0$ can be anything;  \ref{initialize} was just an example.
 I will come back to its interpretation shortly.

\begin{figure}[h!]
\begin{center}
\includegraphics[scale=.3]{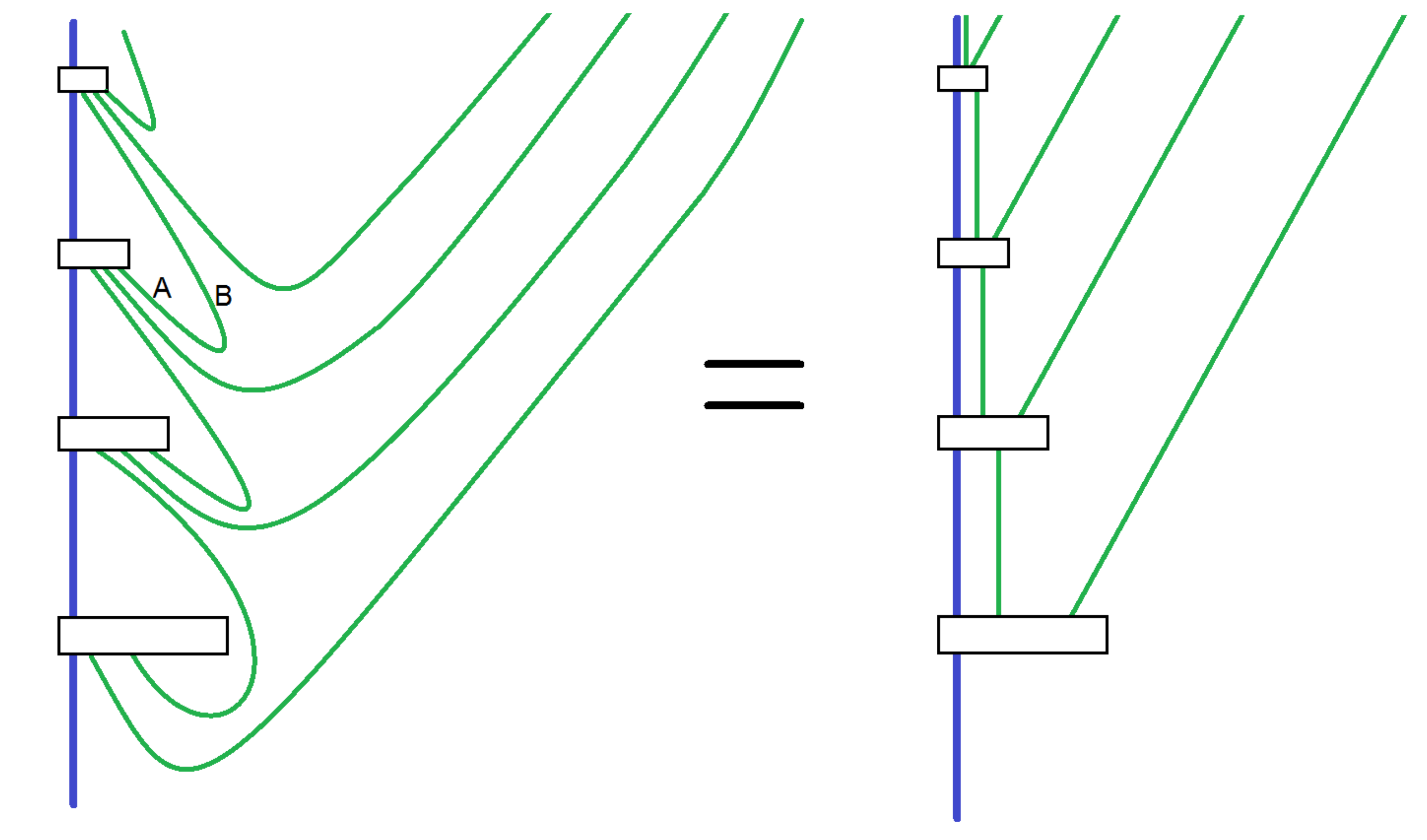}
\caption{The circuit on the left illustrates a simplified history of an evaporating black hole divided into scrambling-time intervals. Every scrambling time new paired $A,B$ qubits are introduced. The old reference qubits fall back to the stretched horizon where they are recycled in the later scrambler. A number of quanta are radiated  to the right at every stage as Hawking radiation. The right figure shows an equivalent diagram with the time-folds undone. }
\label{big-circuit3}
\end{center}
\end{figure}

 The circuit in figure \ref{circuit} is not intended to represent the entire history of the black hole but only one step describing the momentary encoding of the interior in the code. The history from birth to evaporation can be described by the sequence of  scrambling processes shown in figure \ref{big-circuit3}  ---left side.

At any time  the code contains the stretched horizon degrees of freedom (vertical blue lines) and the radiation (slanted lines to the upper right. The zone-modes  and their partners are labeled $B$ and $A.$  For old black holes the code is mostly radiation. The first scrambling process after the collapse process increases the qubit count and may be described by the unitary process in in equations \ref{initial}\ref{scramstate}. But after that the qubit count should not increase which means that the process of figure \ref{toomuch}, with its negative qubits, should be used.

The right side of figure \ref{big-circuit3} shows the diagram redrawn with the timefolds undone so that each box is a square unitary scrambling matrix. From this version it follows that each $B$ mode is almost maximally entangled with the code consisting of the stretched horizon and radiation. The maximal entanglement is not exact but is correct to exponential accuracy. This follows from the properties of scrambled systems \cite{Page:1993df}.

In the earliest scrambling episode the message qubits are counted positively since the qubit count should be increasing rapidly. (Compare with the discussion of young black holes in Section 3.2.)  Once the black hole is scrambled each new message must be counted as having negative qubits using the device of \ref{shift}. In this way the qubit count of zone plus code  can be kept constant. Introducing negative qubits in this way is the origin of the apparent AMPSS contradiction in Section 3.

One may ask why go to all the trouble of introducing a timefold in figure \ref{big-circuit3}? The answer is in order to construct $A$-modes, which once they cross behind the horizon describe the interior. The $A$ modes do not have a life of their own, and we see this especially vividly in the construction involving timefolds.

\subsubsection{The Self-Healing Properties of the Horizon}

\begin{figure}[h!]
\begin{center}
\includegraphics[scale=.3]{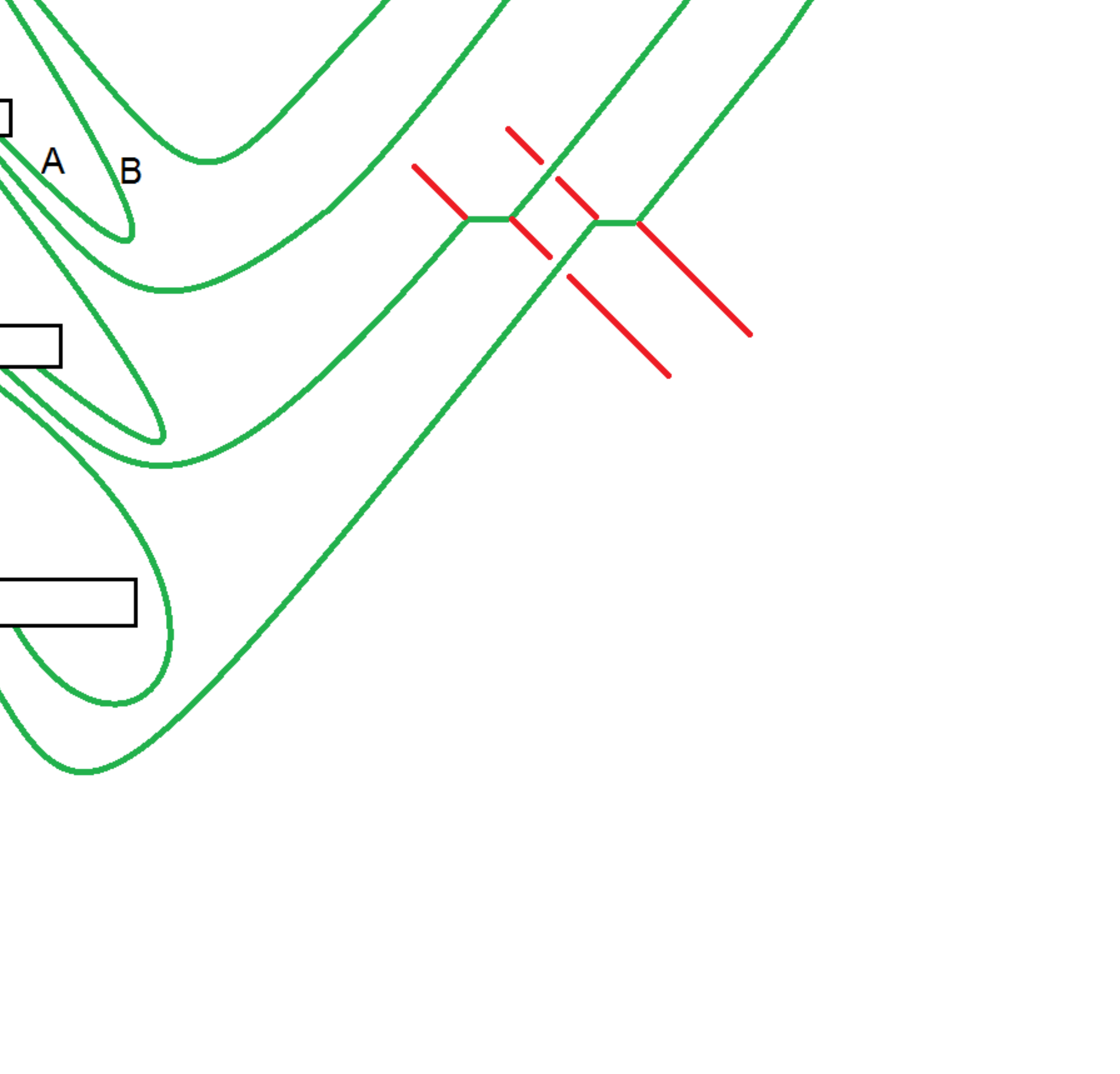}
\caption{ A large fraction of the radiation is suddenly scattered by an environment. The scattered code qubits are not lost; they are  transferred to the environment. The black hole zone remains maximally entangled with the code. }
\label{kink}
\end{center}
\end{figure}

Suppose that some event occurs in which environmental degrees of freedom suddenly interact with  a large fraction of the code qubits. This is  schematically shown in figure \ref{kink}.

\begin{figure}[h!]
\begin{center}
\includegraphics[scale=.1]{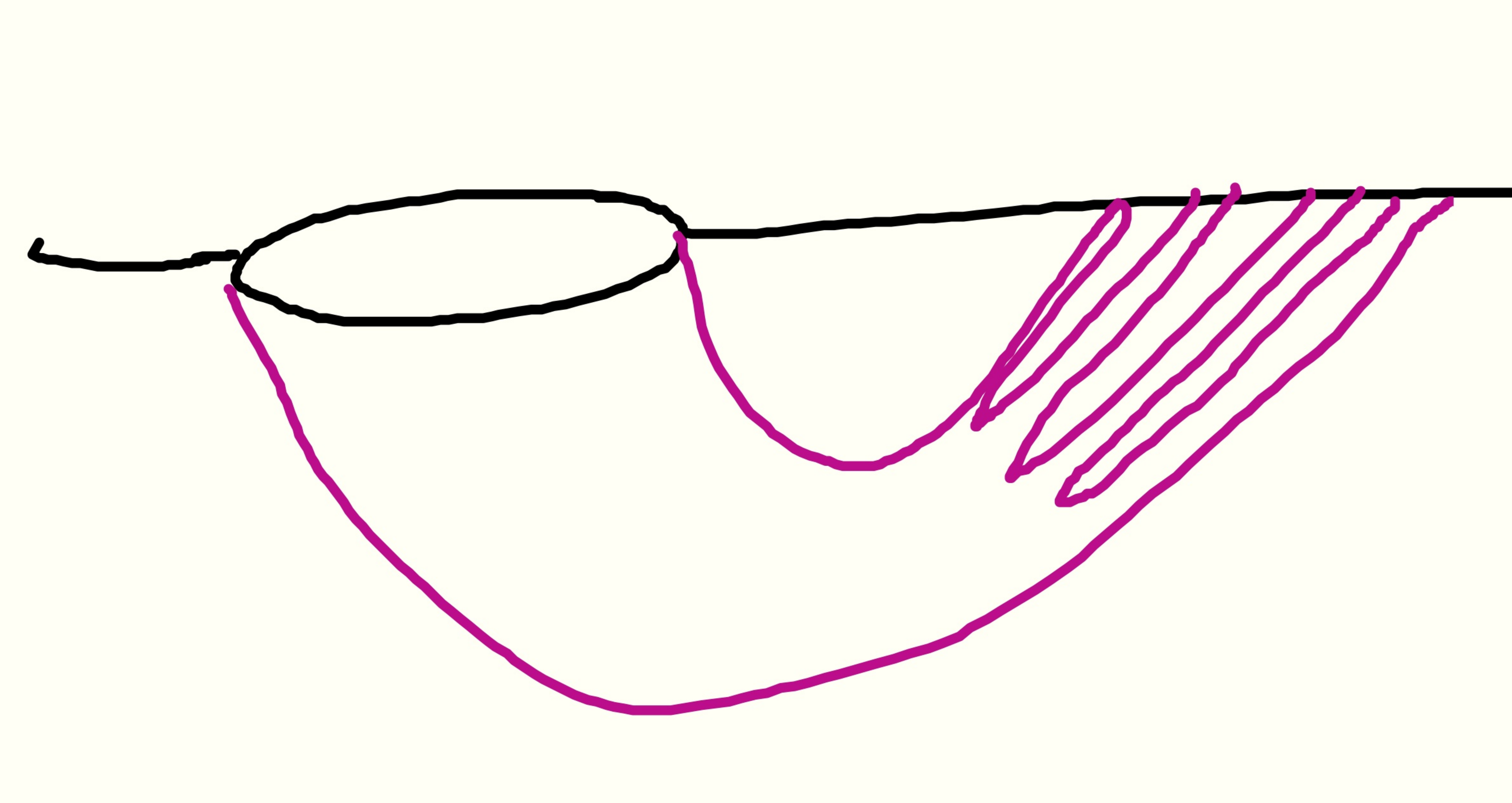}
\caption{ The Einstein-Rosen bridge connecting a black hole to its own Hawking radiation.}
\label{octopus}
\end{center}
\end{figure}

Under certain circumstances this may produce particles in  one or more interior modes by sending signals through
 the Einstein-Rosen bridge connecting the old black hole to its Hawking radiation (figure \ref{octopus}).
In principle  a dangerous firewall could be created for Bob. It is not clear what the criteria are for this to take place, but let's assume that it does happen.
The question is whether   the event  permanently ``scars" the interior of the black hole?

 First of all, note that
there is no way to undo the entanglement between the black hole and the environment without having them come in direct contact.  This means that  there is always a code system available.

 Consider the circuits in figures \ref{big-circuit3}\ref{kink}. Within a scrambling time of the corrupting event a new system of Bell pairs $A,B$ will be introduced. Initially, when the pairs are first introduced the code will not be fault-tolerant since erasing $A$ leaves $B$ with no purification. But the next round of scrambling  will mix $A$ into the
 code and it will become encoded fault-tolerantly.

By this process the fault-tolerant code, and the reference system, will be renewed at each stage. Assuming that this is sufficient for a smooth horizon, we may
 describe this as the ``self-healing" property of the black hole interior.  The time scale for the healing is the scrambling time ( about a millisecond  for a solar mass black hole). Note that contrary to AMPSS, scrambling is not something which destroys the smooth horizon. To the contrary, scrambling continually renews the horizon.

A particular example of a disturbance is a background which creates a  time-dependent Hamiltonian acting on the radiation. If the time dependence takes place on a time scale shorter than the scrambling time, and if it affects more than half the radiation modes, then it may send a temporary firewall to Bob. But if the  dependence is adiabatic
 over a time much greater than the scrambling time, the continual self healing will insure that Bob does not experience any drama as he enters the horizon.

The healing of the interior of the black hole has a very simple conventional explanation. The particles that were created by the disruptive event, and which passed through the Einstein-Rosen bridge,  will fall into the singularity within a scrambling time.

\subsection{Fault-Tolerant Encoding of the Interior}

\begin{figure}[h!]
\begin{center}
\includegraphics[scale=.5]{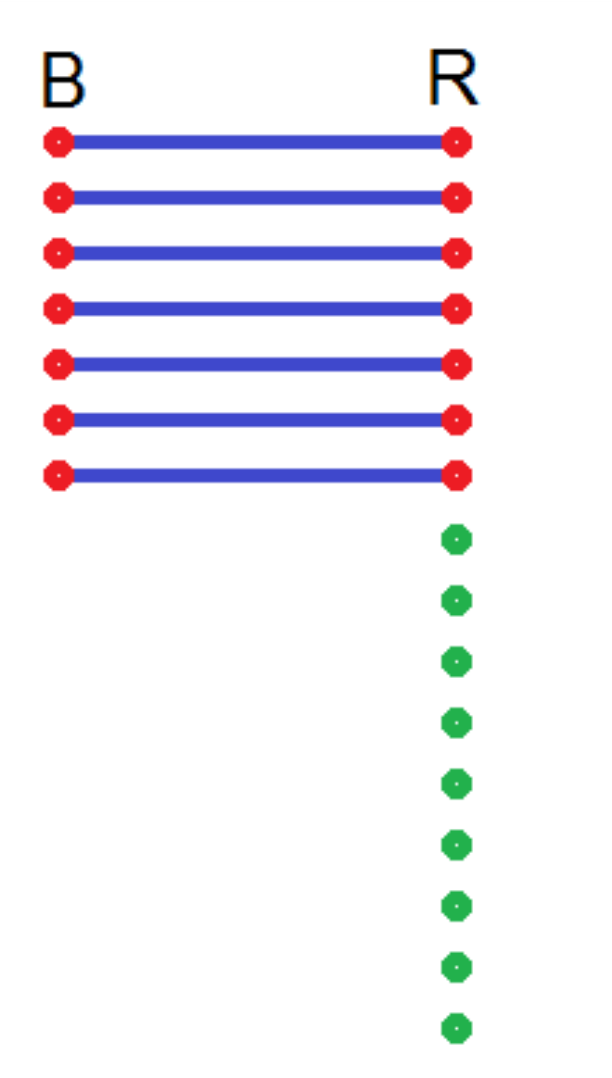}
\caption{ Illustration of $|\Psi_0\ra.$ The red dots connected by blue lines represent Bell pairs and the green dots represent qubits in an initialized state $ {|\n\ra}_0 .$  In the next step the qubits on the right side are scrambled.}
\label{prescramble}
\end{center}
\end{figure}

The state \ref{initial} is schematically illustrated in figure \ref{prescramble}.
It is obviously not fault-tolerant as it is; the loss of any $\a$-qubit is un-recoverable.  But after $U_{scr}$ acts the  code $\c$ becomes fault-tolerant.

 Suppose that before scrambling some agent secretly measured a $1$-qubit operator of one of $k$ message-qubits, thereby disturbing it. There is no way to correct the error without knowing what operator acted, and on which qubit.
However, the situation is very different after scrambling. A corrupted qubit can be detected and corrected. Up to a point this is true for multiple site errors as long as they do not involve too many qubits. When the number exceeds the tolerance limit, the errors become uncorrectable.
The transition from correctable to uncorrectable is sharp and closely related to Page's transition for the information is subsystems \cite{Page:1993df}.

Consider a qubit $B$ in the reference system. Because of the properties of scrambled systems, it is guaranteed that the density matrix of $B$ is maximally mixed and that there is a purification of $B$ in $\c.$
We can think of this purification as the new version of $A.$
Now imagine erasing a particular qubit $C$ from the code. Mathematically that means tracing over $C$ to produce a density matrix for the entire remaining system. Relative to
the message qubit $A,$ the code is fault tolerant for errors that occur on the qubit $C,$ if after erasing $C$ there is still a purification of $B$ in the remaining code system.  Typically after erasing $C$ the purification of $B$ will be exact up to errors exponentially small in the number of qubits of the code.

The special properties of scrambled states insure that a scrambled code is fault tolerant for all $A$ and $C.$
A further generalization is to consider all measurements on a particular set of code qubits $\{C\}.$ Again, the code is fault-tolerant for message-qubit $A$ if, after erasing $\{C\}$ there is a purification of $B.$ For multiple errors (on specified qubits ) a scrambled code is tolerant for all errors on fewer than $\frac{N+k}{2}$ code qubits\footnote{For errors on unspecified qubits the corresponding number is $\frac{N+k}{4}.$ Patrick Hayden, Private communication.}.

\bigskip
It is now possible to state a conjecture relating error correction, and the effect of disturbances of the Hawking radiation on the degrees of freedom behind the horizon.

\bi

\item
Conjecture: \ \

\bigskip
\noindent
A disturbance  on a collection of radiation modes $\{r\}$ will not corrupt a mode $A$ behind the horizon, if the code is fault tolerant, in the sense that tracing over $\{r\}$ leaves a purification of the partner mode $B.$

\ei

\bigskip
\noindent
With this criterion disturbances on fewer than half the radiation qubits will not cause a disturbance to any interior mode $A.$
It follows that any attempt of Alice to send a message to Bob at a time later than the scrambling time must fail unless she acts more than $\frac{N+k}{2}$ radiation modes.

I also argued that if Alice does manage to corrupt the interior it will heal itself over a scrambling time.

\subsection{Partial Scrambling}

Scrambling is not an all-or-nothing business. A chaotic quantum system will scramble in a  time $t_{\ast}$ which for black holes is expected to be logarithmic in the entropy. For shorter times the scrambling will be incomplete.
Partially scrambled codes have varying degrees of fault-tolerance. To my knowledge this has not been studied so what I write here is somewhat conjectural.

We can define partial scrambling by writing the scrambling operator $U_{scr}$ as an exponential,

\be
U_{scr} = e^{-iH_{scr}t_{\ast}}.
\ee

\bigskip
\noindent
where for now, $t_{\ast}$ is just a parameter but later will represent the scrambling time. This allows us to consider partial scramblers

\be
U_{scr} = e^{-iH_{scr}t}.
\ee

\bigskip
\noindent
for $t<t_{\ast}$.

We can define a $t$-dependent numerical measure of the degree of fault-tolerance. If any error involving fewer than $n/2$ qubits of the $(N+k)$-qubit code is correctable ( in the sense that I defined earlier) then we will can say that the code is $n$-tolerant. Roughly speaking it means that as entanglement spreads, the code behaves as if it were scrambled over a fraction $n/(k+N)$ of the qubits.
If the scrambling is complete then the code is tolerant to degree

\be
n= \frac{N+k}{2}.
\label{ntolerance}
\ee

\bigskip
\noindent
If the scrambling is partial the numerical measure of fault-tolerance $n(t)$ will be less than the fully scrambled value. A reasonable expectation is that for $t<t_{\ast}$ the behavior of the fault-tolerance is

\be
n(t) = \frac{N+k}{2} e^{t/t_{\ast}}
\label{growth of n}
\ee

\bigskip
\noindent
For $t>t_{\ast}$ the fault-tolerance will saturate at its maximum value $\frac{N+k}{2}.$
This means that an operator which is correctable for $t \geq t_{\ast}$ may cause a disturbance behind the horizon for much smaller $t.$ For example, even a single qubit operator drawn from the $k$ message-qubits is uncorrectable at $t=0.$

\subsection{State Dependence}

So far I have not discussed the meaning of the state ${|\n\ra}_0$ in equation \ref{initialize}. For simplicity we chose it to be $|00000000....\ra,$ but we could have chosen it to be anything. I also have not commented about the  state of the black hole at the beginning of a scrambling cycle in figure \ref{circuit}. From figure \ref{big-circuit3} we see that these two should be identified.

This brings us to a controversial point; namely the state dependence of the code relating the radiation degrees of freedom to the interior of the black hole. It is very evident that
the way the $k$-qubit message is encoded in $\c$ is extremely sensitive to $ { |\n\ra }_0 .$
In particular we must expect that the operator $R_B$ that encodes data about $A$ will be completely changed if we flip a single qubit in $ { |\n\ra }_0 .$ The implication is that the code relating the CFT to the interior degrees of freedom is highly state-dependent \cite{Verlinde:2012cy}. State dependence and other features of the code has been discussed in detail by Papadodimas and Raju \cite{Papadodimas:2013wnh}.

Such state dependence of the exterior-interior  mapping is not completely new. It occurs in the
simple pull-back---push-forward strategy  for young black holes. Suppose that the
black hole is made by sending in a shell of a particular composition. The shell could be a
coherent electromagnetic wave, a similar gravitational wave, or a mix of the two.
To carry out the pull-back|push-forward procedure, the operator behind the horizon has to be pulled
back to the remote past using the low energy equations of motion in the infalling frame.
Since the equations of motion have to be pulled back through the shell, the operator one
obtains in the remote past will depend on the nature of the shell. That dependence will
also be present after the operator is pushed forward. Therefore the dictionary between
interior operators and operators in the Hawking radiation depends on the initial state
through the dependence on the state of the shell. This is a mild form of the same kind of
non-linear dependence.

\sc
\section{Easy and Hard in the Lab}
\subsection{Strings and Qubits}

The hard-easy distinction, and the criteria for which of Alice's actions send  messages through the Einstein-Rosen bridge, have an analogs in the laboratory model.
One point of difference with the evaporating case is that two sides have the same entropy in  the laboratory model,  corresponding  to being exactly at the Page time. The model does not have the flexibility to go to the limit in which the radiation entropy is much larger than the black hole entropy.

For the evaporating black hole the easy  qubits are the local radiation modes. In the laboratory model the radiation is replaced by Alice's black hole. It is not  obvious how the easy qubits of Alice's black hole should be described, other than to say that they are degrees of freedom of the left-side CFT. In order to illuminate the nature of these qubits I will use a crude model, based on  some old string-theory ideas about black holes  and   Hamiltonian lattice gauge theory.  For definiteness we take the 't Hooft coupling to be  order one and the temperature at the  the Hawking Page transition.

The model is based on the observation that a black hole can be interpreted as a single string in a highly excited state \cite{Susskind:1993ws}\cite{Halyo:1996xe}\cite{Horowitz:1996nw}. If one starts with such a string at weak coupling it forms a spatial random walk, with an energy  proportional to its  total length. The entropy is also proportional to the length, and  counts the ways that the string can ``turn"  after every string-length $\sqrt{\alpha'}. $ For a string of a given entropy, as one increases the string coupling constant
a transition occurs when the  Schwarzschild radius becomes $\sqrt{\alpha'}.$. This defines the crossover or correspondence point.

In context of gauge-gravity duality with 'tHooft $\lambda \sim 1$ the effective strings  are the electric flux tubes which undergo a condensation at the de-confinement transition. In ADS/CFT the role of the  de-confinement transition is played by
the Hawking Page transition. The  transition temperature is of order unity, implying that  the  electric flux tubes are as thick as possible; i.e., the cross section of a single flux tube is of order the radius of the sphere that the gauge theory lives on. It follows that the effective string tension and string length $\sqrt{\alpha'}$ are also order one. Finally, at the Hawking Page transition the black hole horizon area in ADS units is $\sim 1$. What  all this means is that  the Hawking Page transition is right at the correspondence point,  where excited strings transition into black holes. The nature of this transition is discussed in \cite{Horowitz:1996nw}.

One can build a model that reflects these features. Since the black hole is of ADS size, one expects that only the very lowest  approximately homogeneous modes of the gauge theory are relevant. We can regulate ( coarse-grain) the gauge theory by introducing a lattice with only one cubic cell.
The large $N$ Hamiltonian lattice gauge theory \cite{Kogut:1974ag} on a single spatial cube defines a matrix theory that  will roughly have the right properties if the temperature is at the de-confinement transition. At the transition  the cube becomes filled  with a single long electric flux line of length $N^2,$ which covers every link many times. The entropy which stored in the turns and twists of the flux line  is of order $N^2,$ as is the energy.

To see why the theory produces such long strings note that both the energy and entropy of a long string is proportional to its length. This means that the Boltzmann weight has the form

\be
e^{-\beta E + S} = e^{(-c_1 \beta  + c_2)L}
\ee

\bigskip
\noindent
where $c_{1, 2}$ are constants and $L$ is the length of the string. At the transition temperature

$$\beta = \frac{c_2}{c_1}$$

\bigskip
\noindent
the theory is unstable to creating increasingly long strings. By tuning the temperature the length of the dominant string can be set to $N^2.$ These strings wind over gauge theory sphere (the cube) many times.

The effective strings are created by Wilson loops in the lattice model. The simplest Wilson loops are single-plaquette operators. These act locally on the string to move   a few neighboring links.  But from the point of view of the original dual gauge theory they are non-local objects smeared over the entire  sphere. By the UV/IR connection they  should not be identified with the asymptotic boundary, but with the stretched horizon.

We can also consider very long Wilson loops that wind over the sphere (or cube) many times. In particular there are loops of total length $N^2$ which are as complicated as the black hole itself.
For example, suppose we begin with a simple one-plaquette Wilson loop $W,$ and evolve it to construct its precursor $W''$.

\be
W'' = U(t) \ W \  U^{\dag}(t)
\label{looprecursor}
\ee

\bigskip
\noindent
In evolving with $U(t)$ the precursor  $W''$ will grow exponentially rapidly into a superposition of  long Wilson loops of length $N^2.$

It is possible to work with the gauge theory degrees of freedom but it is simpler to just go to a discrete string description with $N^2 $ sites along the string. The details of how a lattice string is described \cite{Klebanov:1988ba} are interesting but not very important. All we need to know is that it is described by a long string of discrete variables similar to qubits. The string can clearly be used as a code system.

The easy-hard distinction is fairly clear:  operators which act on short lengths of string are easy, and operators which act on long lengths,  comparable to  half  or more of the total length are hard.

 For the two-sided laboratory model we introduce  two independent lattice gauge cubes that begin in a thermofield-double state. To visualize the state consider two identical long strings of length $N^2,$ one on each cube. Begin with a state in which the two strings are mirror images of each another. There are of course many such states.
Next construct a linear superposition of all such mirror image states. That is the analog of the thermofield-double state.

From there the system evolves as two independent decoupled lattice systems. In the dominant configuration the total length of string on either side continues to be $N^2.$

The thermofield state is not stationary; with time it evolves  in a manner described by Hartman and Maldacena \cite{Hartman}\cite{Maldacena:2013xja}.
The evolution can be modeled by a system of $2N$ qubits shared between left (Alice ) and right (Bob).
 The initial thermofield-double state is entangled but unscrambled. It closely resembles a system of $N$ Bell pairs shared between the two sides,
\be
|TFD\ra = (|00\ra+|11\ra^{\otimes N}
\ee

\begin{figure}[h!]
\begin{center}
\includegraphics[scale=.5]{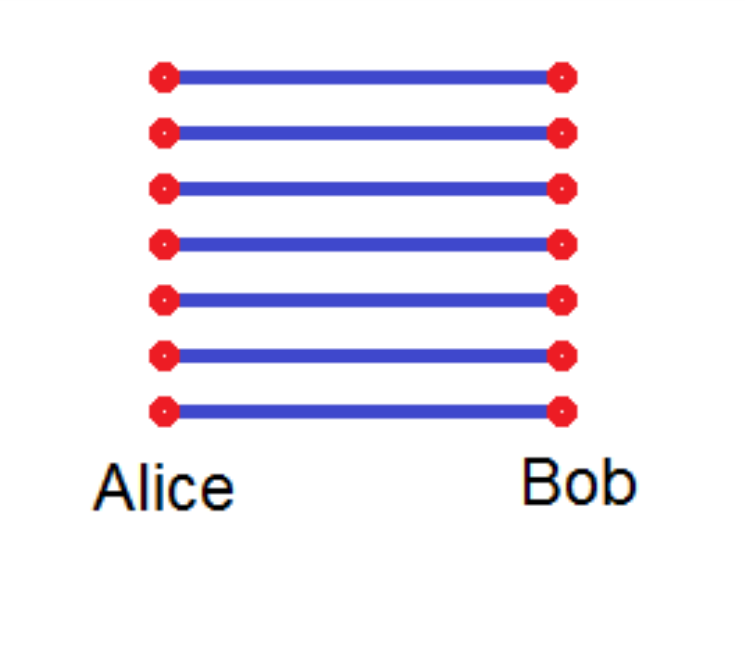}
\caption{ The thermofield-double state is modeled as a product system of $N$ Bell pairs. }
\label{TDF}
\end{center}
\end{figure}

\bigskip
\noindent
It is illustrated in figure \ref{TDF}.
The state is similar to \ref{initial} except that there are no additional qubits to represent the initial state of the black hole.

The thermofield state is maximally ``horizontally entangled,"  but there is no ``vertical entanglement" at all\footnote{Horizontal and vertical refer to the direction in figure \ref{TDF}.}.
Thought of as a code, it is not fault-tolerant. If a qubit on the left is accidentally flipped there is no way to recover it by looking at the remaining qubits. If it is erased then a corresponding qubit on the right has no purification.

Hartman and Maldacena show that as time progresses the two-sided system begins to develop vertical entanglement.
This happens as a consequence of separate evolution on the two sides.
The evolution is described by unitary operators $U_{L,R}(t),$

\be
|TFD\ra \to U_L(t) U_R(t) \left\{|00\ra+|11\ra^{\otimes N} \right\}
\label{partscramb}
\ee

\bigskip
\noindent
The $U_{L,R}$ operators are fast-scramblers \footnote{The logarithmic scrambling time is correct for black holes of ADS scale whose entropy is of order $N^2.$ For larger black holes the scrambling time has a more complicated form but the general  conclusions are the same.}. The result, as shown in \cite{Hartman}, is a massive increase in vertical entanglement until it saturates in a fully scrambled state. Although Hartman and Maldacena did not describe it in terms of fault-tolerance, this growth of vertical entanglement means that the code goes from being fragile to being maximally
fault-tolerant; and stays fault-tolerant at least for times less than the quantum recurrence time.

All of these observations also apply to the two-cube model.

There are gauge theory operators that act on the string  in familiar ways. Consider a single plaquette Wilson loop. The effect of the Wilson loop is to vary the configuration locally. When it acts it does so on a small number of links. In other words it acts as an easy operator.

On the other hand one can consider Wilson loops of very long extent. A Wilson loop of length $N^2$ has a massive effect on all of the links. It certainly is  a hard operator.

\begin{figure}[h!]
\begin{center}
\includegraphics[scale=.6]{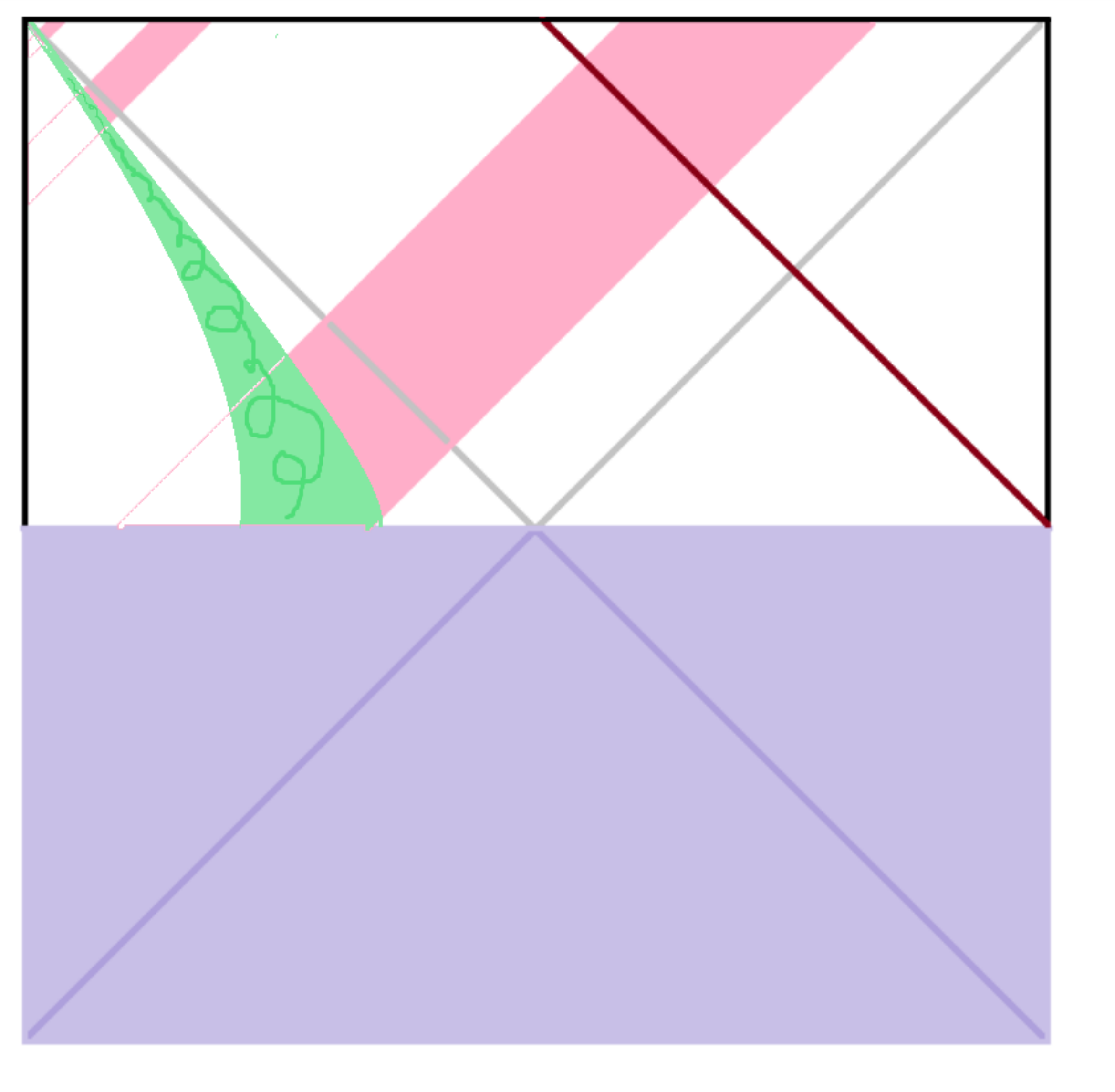}
\caption{ The brown line represents Bob jumping into his black hole at $t=0.$ The green curve represents  the stretched horizon where the entropy is stored in the form of a long string.
The various pink regions are Alice's attempts to send messages to Bob by acting with a single qubit at various times. If she send the message from $t=0$ Bob just  receives it before hitting the singularity. If she sends much later Bob cannot receive it. }
\label{correctable}
\end{center}
\end{figure}

Let's now return to Alice's attempt to communicate with infalling Bob.
The earliest that Bob can jump into his black hole, and the earliest that Alice can send a message, is $t=0.$ At that time the state is the thermofield state built of mirror image strings.
Alice can apply a single qubit operator---in other words a short Wilson loop which covers the gauge theory sphere once. If we think of that as inducing an error, it is analogous to a one qubit error. Tracing over that qubit would destroy the purification of a corresponding qubit on Bob's side,
 thereby potentially corrupting the partner behind the horizon.
 Schematically   this is illustrated in figure \ref{correctable}.

From the figure one sees that if Bob jumps in as early as possible he can  receive Alice's message. Therefore the action of the simple qubit operator acting at $t=0,$ must be un-correctable. Erasing it must erase the purification of one $B$-mode which is indeed true for the product of Bell pairs. In the string picture Alice has acted on one string loop at the stretched horizon.

Now let's increase the time at which Alice acts and Bob jumps, to $t$ with $ 0< t << t_{\ast}.$
The two gauge theories will  partially scramble as in equation \ref{partscramb}. As a result the code will become more fault-tolerant. Very shortly, in a time of order the inverse temperature, $n(t)$ will become  $>>1.$ At that time all single qubit errors will be correctable and Alice will no longer be able to get a message to Bob by acting with a simple Wilson loop.

On the other hand, she can apply a precursor of a one plaquette Wilson loop \ref{looprecursor}.   For the range $ 0< t << t_{\ast}$  the partially scrambled precursor of a plaquette operator is a more complex non-local operator than the single plaquette itself; although not nearly as complex and non-local as it will be at the scrambling time. In terms of Wilson loops the length of the precursor-loop will quickly grow with $t.$

Evidently, the more complex the precursor, the deeper it reaches into the bulk.  The error generated by acting with the precursor at time $t$ will be less easy to correct, than single qubit operator.

Finally at the scrambling time $t_{\ast}$
the code will be maximally fault-tolerant,  but the precursor  will be extremely complicated. It will contain order S qubits (links) in the gauge theory and will be  un-correctable. Nevertheless it is extremely fine-tuned so that when it acts it only corrupts a single qubit behind Bob's horizon.
Obviously the action of the short Wilson loops is similar to  easy operators, and the tremendously complicated precursors-loops are like hard operators.

From this point of view the challenge  AMPSS poses is to understand why the increasing complexity of the precursor  allows it to reach so deeply into the bulk that it can send a message to $A,$ despite the fact that that the  short Wilson loops cannot.  Although this challenge has not been met in this paper,  it appears to be exactly the same  challenge as understanding how precursors differ from simple Wilson loops \cite{Giddings:2001pt}\cite{Freivogel:2002ex}. To my knowledge this is  not fully understood yet.

Finally, does the laboratory shed any light on the relation between signals and error correction? Perhaps just a bit.
First let's review the evidence that applying $A''$ does cause a signal to propagate to $A.$ The argument was:

\bigskip

1) Applying $A'$ has the effect of sending such a signal. The reason is dynamical. $A'$ is in the causal past of $A$ and any ordinary dynamical considerations would say that applying it should create a signal.

\bigskip

2) $A''$ was constructed to have the same effect as $A'.$

\bigskip

The scrambled qubit $A''$ is a subsystem of the left CFT and can be traced over. Doing so has exactly the same effect as tracing over $A'.$ But $A'$ and $B$ are entangled pairs in the Thermofield-double state. In other words erasing $A'$ destroys the purification of $B.$ It follows that erasing $A''$  also destroys the purification of $B.$ Thus we see that at least in this case dynamics and the criterion conjectured in Section 4.4 agree.

\sc
\section{Conclusion}
In this paper I've made  no attempt to prove that firewalls are absent in all circumstances. Indeed ER=EPR raises the possibility that an angry Alice can hit Bob with a nasty shockwave as he crosses the horizon \cite{Shenker:2013pqa}\cite{VanRaamsdonk:2010pw}.
What I have  assumed is that firewalls are not inevitable--- particularly so if the black hole begins  with a smooth horizon---and then asked what new concepts are required to resolve the various paradoxes. In a sense I am trying to turn the firewall inevitability arguments  into arguments for new physical concepts  needed to reconcile unitarity and complementarity.

It seems that this will require new physics behind the horizon, one example being Alice's ability to send a message to Bob which he can receive just after passing the horizon\footnote{It  is possible that in the real world---as opposed to the laboratory model---Harlow-Hayden complexity \cite{Harlow:2013tf} would never allow Alice to carry out the procedures needed to send the message. If anything, that would make the argument for firewalls less compelling since the basic AMPSS experiment would not be possible.}. Alice can be arbitrarily far away, and Bob can receive the signal immediately after crossing the horizon.  All that Alice needs is a system which is sufficiently entangled with Bob's black hole, and a quantum computer that can carry out very complicated quantum operations \cite{Maldacena:2013xja}.

This is a twist on two commonly held incorrect sci-fi ideas; the first being that superluminal signals can be sent through wormholes; and the second that superluminal signals can be sent using entanglement. ER=EPR does not allow superluminal signals, but it gets very close, in the sense that there is no limit on how soon after horizon crossing Bob can receive Alice's message.

ER=EPR, timefolds, precursors and fault-tolerant codes---are new ingredients. They may seem very radical but they were introduced for the conservative purpose of reconciling unitary evolution with the equivalence principle. Quoting Polchinski, they are ``either very crazy or very deep."

These ingredients were used to examine various arguments for a breakdown of black hole complementarity. The original AMPS argument depended on the hidden ``proximity" postulate. Violating the proximity postulate requires that very distant manipulations of the Hawking radiation can somehow produce particles just behind the horizon---an apparently wildly non-local effect. That is where ER=EPR comes in; the produced particles don't travel long distances through the external space but arrive by passing through the Einstein-Rosen bridge. In this way a particle can materialize arbitrarily close to the horizon, but always on the inside. The bridge is non-traversable but just barely.

To send such particles after the bridge has grown long, operators which generate them must behave like precursors. In effect Alice must turn the clock back at her end, and send the particle from an earlier time. The mathematical description of such an operation is the timefold. Sending a particle in this manner would contradict locality if the particle could arrive outside the horizon at Bob's end, but that never happens.

Precursors lead to some strange ambiguities about whether Alice's measurement of $A''$ (or $R_B$) is actually detected  by Bob when he jumps in. If one is to believe the analysis, then by measuring the precursor $A'',$ Alice creates two histories, one in which Bob detects a particle behind the horizon, and one in which he does not. But if after making the measurement, Alice herself jumps in to check, then there will  be only the history containing the particle. Timefolds provide an way to visualize these ambiguities.

Next we turned to  arguments of AMPSS \cite{Almheiri:2013hfa}\cite{Joe}  concerning whether or not the interior of a one-sided  ADS black hole can be encoded in the dual CFT degrees of freedom. There is an apparent contradiction, that was exposed by AMPSS, between the properties of the interior modes $A$  and the counting of states of black hole states. The argument leads to the paradoxical conclusion that creation operators must lower the entropy and annihilate half the states. One could take this to be an indication that the region behind the horizon cannot exist.

However similar behavior is known in ordinary gauge theories. It is characteristic of non-gauge invariant operators which cannot be realized as  acting on the physical space of states. Only gauge invariant modifications of the operators make sense as physical quantities. In the gravitational case the gauge invariance is diffeomorphism invariance and it seems likely that invariant versions of the operators can be realized in the dual CFT. These gauge invariant versions describe not only the modes behind the horizon, but also the clocks, rods, and apparatuses needed to measure them.

However this does not mean that someone outside the black hole can make a measurement that can tell what happens behind the horizon. The CFT encodes the interior but only in the same way that the past encodes the future. The interior of the black hole should be thought of as being in the future of everything directly  described by the  gauge theory. The CFT is a machine for constructing  initial states on the horizon for later events behind the horizon. The geometry of the region described by the CFT is not geodesically complete. Gauge-theory time $t=\infty$  should be identified with the horizon.

The AMPSS commutator argument, and outgrowths of it, appear to be the most difficult to answer. The current situation  is that there are no clear criteria for what operations by Alice will send signals through the Einstein-Rosen bridge to Bob.  \cite{Maldacena:2013xja} conjectured a connection a criterion based on thinking of the interior as a fault-tolerant message encoded in the Hawking radiation. According to this idea actions of Alice can be considered to create errors. Correctable errors do not send signals and un-correctable errors do. Correctable errors correspond to easily measured operators such as local radiation modes. Hard-to-measure operators such as precursors correspond to uncorrectable errors.
While something like this may be true, a precise version of the  conjecture seems elusive.

The message interpretation of the interior modes requires a somewhat unconventional interpretation of these modes as carrying a negative number of qubits. The fact that  the interior modes carry negative energy was the basis for the first AMPSS paradox, but the interior modes must also carry negative information. The concept of negative qubits was explained as another manifestation of timefolds. Of course negative information has no real meaning and adds to the feeling that  $A$ has no real life of its own.

One of the main points of disagreement between the conclusions of AMPS and of this paper has to do with the role of scrambling. According to AMPS the effect of scrambling is to destroy the smooth horizon of a young black hole. The arguments of Section 4.31 lead to the opposite conclusion: scrambling leads to continual renewal of the smooth horizon. Should an event occur that sends a dangerous message through the Einstein-Rosen bridge, the resulting scar will be healed in a scrambling time.

  Finally in Section 6 a crude model for ADS black holes as single strings was used to illuminate the meaning of easy and hard operators in the laboratory model. Wilson loops that wind a small number of times around the gauge theory sphere are easy: long Wilson loops of length $N^2$ are hard. Precursors are examples of such hard operators. There is evidently a close relationship between the puzzle of easy and hard operators, and the somewhat mysterious nature of precursors.  Understanding  precursors may help illuminate the easy-hard distinction.

\bigskip

\bigskip

Evidently the study of the interiors of black holes is an extremely rich and strange one that is bound to be full of surprises.

\sc
\section{Added Comment on Non-linearity}

\bf

There\footnote{This section was originally included but then removed in order to shorten the paper. In light of some recent discussions concerning (non)linearity \cite{Papadodimas:2013jku}\cite{Papadodimas:2013wnh} I decided to put it back in version 2.} are a number of arguments that revolve around the assumption that a firewall-operator can distinguish states with a firewall from states with a smooth horizon \cite{Almheiri:2013hfa}\cite{Bousso:2013wia}\cite{Marolf:2013dba}. The idea is simplest when applied to a one-sided black hole.
The arguments can be made mode-by-mode. To say that Bob hits a firewall in the mode $A$ simply means that he encounters a particle in that mode. If enough modes are occupied then the firewall becomes a potent obstruction for Bob.

Let $F$ be a firewall operator for a single $A,B$ mode-pair that takes the value $0$ if there is no local firewall, and is positive if there is a firewall. $F$ is assumed to act on the degrees of freedom of a one-sided black hole and therefore should be an operator in the one-sided physical Hilbert space.

We can consider a state in which the zone degree of freedom $B$ is unentangled, i.e., it is in a pure state. Then, according to hypothesis there will be a large probability $\sim 1$ for there to be a local firewall (a particle) at the site of $A.$ In other words $\la F\ra \sim 1$ for such states.
The argument uses the fact that the unentangled states are complete; i.e., they span the Hilbert space. If one now expands the thermal ensemble in a basis of such states one will find that mode-by-mode $\la F\ra \sim 1.$ But if every mode has a probability of order unity to contain a firewall particle, there will certainly be a large number of firewall particles in the thermal ensemble of a one-sided black hole.

As an example of a firewall operator \cite{Almheiri:2013hfa} uses the number operator associated with the linear combination $(A+B)$ which straddles the horizon. If we assume that $A$ is some gauge invariant version of the modes behind the horizon then it should be represented in the right-sided CFT space of operators.
But in that case its expectation value would be the same for any two-sided state which has the thermal density matrix for the right side---independently of what purifies it.

There is something wrong with this kind of firewall-operator argument because the purification could be the thermofield-double state\footnote{This point made by J. Maldacena. }. The argument is so general that it would imply a
positive value for the firewall operator in the thermofield-double state, but that's the one state where we have strong reasons to believe that there is no firewall.

There are at least two possible reasons why the firewall operator argument may be wrong. First of all it is based on the properties of a non-gauge-invariant operator $A$ which does not exist in the physical Hilbert space. If $A$ is made gauge invariant then it should be represented in the one-sided CFT and Maldacena's argument should apply. But there is possibly a deeper point. The argument implicitly assumes that the existence of a firewall is a conventional observable in whatever quantum system describes the black hole. That would imply that the
class of states with/without firewalls form a linear subspace. Suppose we change the wording from: 
 
\bn
 Is there a firewall?  

\bn to:  

\bn
Is there an Einstein-Rosen bridge?  

\bn
The ER=EPR principle suggests that the answer is not governed by a linear operator, but by a non-linear entanglement criterion. To quote Polchinski, "Entanglement matters."

This is not the first time that black hole complementarity has required an element of non-linearity in the global description of quantum states. Quantum cloning  would require a non-linear evolution of states. An allowable classical  cloning operation,

\be
a |0\ra + b|1\ra  \ \to \ a|00\ra + b|11\ra
\ee

\bn
is linear and leads to an entangled state. The cloning operation,

\be
a |0\ra + b|1\ra  \ \to \    (a|0\ra + b|1\ra) (a|0\ra + b|1\ra)
\ee

\bn
is non-linear and leads to a product state, but it violates the linear rules of quantum mechanics. The original information-loss argument of Hawking assumed that information could not be cloned, in the sense that it could not be both radiated in Hawking radiation and transmitted through the horizon.

Simply stated, complementarity argued that cloning is OK as long as no one can see it. To put it another way, the linear operational rules of quantum mechanics must be respected for ordinary observations  within any causal patch. In the cloning experiment there are three causal patches. The first is Charlie's patch---the exterior of the black hole---that contains the Hawking radiation. The second is Bob's patch. Bob falls through the horizon after collecting a radiated qubit. The qubit is a clone of one that Alice carried in earlier. The third patch is Alice's. If one tries to construct a global description that contains  Charlie, Alice, and Bob, the description will contain two copies of the same qubit.

The original claim of complementarity was that conventional linear quantum mechanics is the description of a single causal patch. The non-linearity of cloning is part of putting patches together into a global description, but it is not part of the operational description in any single causal patch. The argument for how observation of cloning is precluded is well known and I won't repeat it here.

It seems to me that the significance of AMPS(S) is that it gives us an additional  example of how non-linearity is required for a global description, when multiple causal patches are combined into a global description. According to complementarity, that's OK, as long as no single patch requires a non-linear description. The challenge posed by AMPS(S) is to show that while the formal reconstruction of the interior requires non-linearity, observations within a single causal patch do not.

The viewpoint that the reconstruction of a global description from many local causal patches will require  non-linear generalizations of quantum mechanics has obvious consequences for cosmological space-times such as de Sitter space, and especially, eternal inflation.

\rm

\section*{Acknowledgements}

Much of the material in this paper represents conversations that I had at the Fire or Fuzz meeting at KITP in August 2013.

I am especially grateful to Juan Maldacena and Daniel Harlow for extended discussions in which the  firewall arguments were identified, and counter arguments proposed; to Stefan Leichenauer for explaining timefolds to me; Steve Shenker and Douglas Stanford for discussions about generic states of two-sided black holes; and Patrick Hayden for clarifying some concepts of error correction.

Support for this research came through NSF grant Phy-1316699 and the Stanford Institute for Theoretical Physics.

\end{document}